\documentclass[aps,prd,floats,floatfix, twocolumn,amssymb,amsmath,
superscriptaddress,nofootinbib,showpacs,longbibliography]{revtex4-2}

\usepackage[T1]{fontenc}
\usepackage[utf8]{inputenc}
\usepackage{txfonts} 
\usepackage[normalem]{ulem} 

\usepackage{graphicx}
\usepackage[linktocpage,breaklinks]{hyperref}
\usepackage[capitalize]{cleveref}
\usepackage[usenames,dvipsnames]{xcolor}
\hypersetup{colorlinks=true,citecolor=NavyBlue,
linkcolor=NavyBlue,urlcolor=NavyBlue}

\usepackage{multirow,array}
\DeclareMathAlphabet{\pazocal}{OMS}{zplm}{m}{n}
\usepackage{microtype}
\usepackage{booktabs}
\usepackage{subfigure}

\usepackage{journals}

\newcommand{\uiuc}{\affiliation{Department of Physics and Illinois Center for Advanced Studies of the Universe,\\University of Illinois Urbana-Champaign, Urbana, Illinois 61801, USA}}

\begin{document}

\title{Hybrid-star asteroseismology across sharp phase interfaces: \\ Slow and rapid conversion in the $f$- and $p_1$-mode spectra}

\date{\today}

\author{Zoey Zhiyuan Dong}
\email{zd26@illinois.edu}
\uiuc

\author{Hao-Jui Kuan}
\email{hjkuan@illinois.edu}
\uiuc

\author{Nicolas Yunes}
\email{nyunes@illinois.edu}
\uiuc

\begin{abstract}

First-order phase transitions in neutron-star interiors can leave
signatures in the polar quasi-normal-mode spectrum. We compute the
complex quadrupolar $f$- and $p_1$-mode spectra in full general relativity for cold,
nonrotating hybrid stars with a sharp hadron--quark interface. We model the equation of state by
matching a hadronic SLy4 phase to a constant-sound-speed quark phase. We compare slow phase conversion, in which fluid elements
retain their phase and the interface moves with the fluid, with rapid
phase conversion, in which matter changes phase as it crosses the boundary
and the interface follows phase equilibrium rather than the fluid,
together with an algebraic interpolation between these limits. On
identical equilibrium backgrounds, the two physical limits produce a
sizable separation in the real part of the $p_1$-mode frequency. This $p_1$ frequency
separation is largest when the local tangential-displacement fraction
of the eigenfunction changes rapidly across the interface. 
Across the surveyed equation of state parameter space, rapid conversion generally lowers the $p_1$-mode frequency relative to slow conversion, with compactness-averaged separations reaching $\sim3.5\,{\rm kHz}$ on disconnected twin-star branches. The largest separations occur at relatively low transition pressures and large density discontinuities.
By contrast, the $f$-mode is only
weakly affected by the interface condition for individual stellar
models. Across the sampled constant-sound-speed parameter space, departures exceeding
the hadronic SLy4 reference residual occur mainly at low transition
pressure, large density discontinuity, or both, with different
boundaries for the real- and imaginary-part fitting variables. Thus,
the $p_1$-mode probes the local interface response, whereas the
tracked $f$-mode reflects the global departure of the hybrid sequence
from the hadronic calibration.

\end{abstract}
\maketitle

\section{Introduction}

The detection of gravitational waves (GWs) from compact binaries has opened a new observational window into the internal structure of neutron stars~\cite{LIGOScientific:2018cki}. In binary-neutron-star systems, the inspiral signal probes the bulk tidal response of the stars, for example through the tidal deformabilities~\cite{Flanagan:2007ix,Hinderer:2007mb}. The postmerger phase, when detected, will contain additional information about the oscillation spectrum of the merger remnant~\cite{Stergioulas:2011gd,Bauswein:2011tp,Takami:2014zpa}. After the merger, the remnant is a highly dynamical object that can undergo strong nonradial oscillations and emit gravitational radiation at characteristic frequencies. The dominant spectral features of the postmerger GW signal are commonly interpreted as arising from fluid oscillations of the remnant, with frequencies determined by the remnant mass, compactness, rotation, thermal state, and dense-matter equation of state (EOS)~\cite{Stergioulas:2011gd,Bauswein:2011tp,Takami:2014zpa,Lioutas:2021jbl}. Therefore, studying stellar oscillation modes provides a useful way to connect GW spectral features to the internal properties of neutron-star matter.

The oscillation spectrum of a compact star is described in terms of quasi-normal modes, whose complex frequencies encode both oscillation (real part) and damping rates (imaginary part). Among the fluid modes, the fundamental or $f$-mode (describing a global, nonradial surface-gravity oscillation) has received particular attention because it couples efficiently to gravitational radiation. Early studies established empirical relations between its frequency and damping time and the stellar mean density or compactness \cite{Andersson:1997rn,Benhar:2004xg,Tsui:2004qd}. Subsequent work introduced relations based on an effective compactness, and hence the moment of inertia, and systematically examined their validity \cite{Lau:2009bu,Chirenti:2015dda}. The GW damping-time relations were further investigated in~\cite{Lioutas:2020vzi}, while separate EOS-insensitive relations involving the tidal deformability were developed in~\cite{Chan:2014kua,Sotani:2021kiw}. 
More recent studies have tested these relations for nonstandard high-density structure, including self-bound quark stars \cite{VasquezFlores:2017uor,Zhao:2022tcw}, constant-sound-speed (CSS) hybrid stars and extended slow-stable branches \cite{Zhao:2022tcw,Ranea-Sandoval:2023ixr}, Gibbs mixed phases \cite{Zhou:2023nzm,Kumar:2023ojk,Podder:2025qaz}, and pasta or twin-star constructions \cite{Pradhan:2023zmg,Zheng:2024tjl}. Full-general relativity (GR) calculations have also considered composition-rich hybrid models containing hyperons and delta baryons \cite{Rather:2024mtd}. Approximate universality survives within several of these samples, whereas systematic departures arise in some broader exotic-composition samples.
Pressure modes, such as the $p_1$-mode, are less universal than the $f$-mode because their frequencies depend more strongly on the internal acoustic structure and composition of the star \cite{Kokkotas:1999bd,Kunjipurayil:2022zah,Sotani:2021kiw,Kumar:2023ojk}. This sensitivity extends to phase-transition-induced changes in the sound-speed profile and to sharp density discontinuities \cite{Miniutti:2002bh,Sotani:2010mx,Ranea-Sandoval:2019miz,Thakur:2024ijp,Laskos-Patkos:2024otk}.

Although nuclear-physics input~\cite{Huth:2021bsp, Reed:2021nqk}, pulsar mass measurements~\cite{NANOGrav:2019jur}, NICER radius measurements~\cite{Miller:2019cac, Miller:2021qha, Miller:2025qfq, Riley:2021pdl, Riley:2019yda}, and GW observations~\cite{LIGOScientific:2018cki,De:2018uhw,LIGOScientific:2020aai} have placed important constraints on the neutron-star EOS, these constraints primarily restrict bulk properties of the star, such as masses, radii, and tidal deformabilities~\cite{Xie:2020rwg,Dietrich:2020efo,Yunes:2022ldq,Lattimer:2021emm,Essick:2023fso,Chatziioannou:2024jsr}.
These quantities do not uniquely determine the microscopic composition of matter at the highest densities reached in the inner core, because different high-density EOS models can produce similar macroscopic observables and inspiral-based constraints remain affected by EOS degeneracies and measurement uncertainties~\cite{Zhu:2025dea}.
One important possibility is a first-order phase transition from hadronic matter to deconfined quark matter~\cite{Alford:2015gna,Xie:2020rwg,Essick:2023fso}.
Such a transition can produce a hybrid star with a quark core separated from the hadronic envelope by a sharp density discontinuity.
The transition is usually parametrized by a transition pressure $P_t$ (which fixes the pressure level at which the interface appears in a stellar model) and a density jump $\Delta\varepsilon$ (which quantifies the strength of the discontinuity)~\cite{Alford:2015gna,Essick:2023fso}.

A sharp density discontinuity can also support a finite family of \textit{discontinuity $g$-modes}, sometimes also called interface or $i$-modes or sharp-interface $g$-modes~\cite{Finn:1987zoj,Sotani:2001bb, Miniutti:2002bh}. The restoring buoyancy of these modes is localized at the density jump, and they are to be distinguished from composition-driven core $g$-modes. The latter arise in extended mixed phases and in smooth crossovers, where buoyancy is distributed through a finite region of the star and arises from the difference between the equilibrium and adiabatic sound speeds \cite{Wei:2018tts,Jaikumar:2021jbw,Constantinou:2021hba,Zhao:2022toc}. 
Unlike core $g$-modes, discontinuity $g$-modes depend on how rapidly matter converts from one phase to the other, when stellar oscillations displace fluid elements across the phase boundary.
In the \textit{slow-conversion limit}, the conversion timescale is long compared with the oscillation period, so fluid elements retain their original phase and the interface moves with the fluid. In the \textit{rapid-conversion limit}, the conversion timescale is short compared with the oscillation period, so fluid elements reaching the interface promptly convert into the new phase.
The interface therefore does not comove with the fluid; instead, its instantaneous location is determined by phase equilibrium.

But a sharp density discontinuity not only gives rise to a new finite family of discontinuity $g$-modes, but it also affects all other modes in the star, including $f$- and $p$-modes.
Existing calculations of conversion-dependent nonradial oscillations have focused mainly on mode sectors other than $p_1$. The full-GR junction conditions for polar perturbations in the slow- and rapid-conversion limits were derived in~\cite{Tonetto:2020bie}, which also showed that the discontinuity $g$-mode has a finite frequency in the slow-conversion limit but collapses to zero frequency in the rapid-conversion limit.
Subsequent Cowling and full-GR calculations studied $f$  and discontinuity-$g$-mode spectra for sequential-transition and magnetized hybrid-star sequences, whose stable portions depend on the assumed conversion regime \cite{Rodriguez:2020fhf,Mariani:2022xek}. Discontinuity-$g$-mode relations have also been obtained for slow-stable twin-star branches in the Cowling approximation \cite{Rodriguez:2025oes}, while complex axial $w_I$-mode relations have been studied over rapid- and slow-conversion stellar samples \cite{Ranea-Sandoval:2022bit}.
These calculations show that the conversion regime affects both stellar stability and several parts of the oscillation spectrum, but none isolates its effect on the complex $p_1$-mode by applying both physical polar junction conditions to the same equilibrium star. 
Conversely, previous complex full-GR $p_1$-modes calculations for stars containing sharp density discontinuities or first-order transitions \cite{Miniutti:2002bh,Thakur:2024ijp} employed a single interface prescription. To our knowledge, no previous calculation has combined these two ingredients in full GR by comparing the complex quadrupolar $p_1$ spectrum under both physical conversion conditions on identical sharp-interface backgrounds, and then relating the resulting spectral separation quantitatively to the local eigenfunction structure at the interface.

In this paper, we address this missing $p_1$ problem and use the $f$-mode to investigate a complementary question about the global hybrid-star structure. In particular, we compute the complex quadrupolar $f$- and $p_1$-mode spectra in full GR for cold, nonrotating, nonmagnetized hybrid stars with a sharp hadron--quark interface. We model the EOS by matching the SLy4 hadronic EOS to a CSS quark-matter EOS at a transition pressure $P_t$. Keeping the SLy4 baseline and the quark-matter sound speed fixed, we vary the transition pressure $P_t$ and the energy-density discontinuity $\Delta\varepsilon$ to construct a two-dimensional family of hybrid EOSs. 
For each member of this EOS family, we then generate a sequence of equilibrium stellar models by varying the central pressure, and then we compute the mode spectrum using the slow- and rapid-conversion junction conditions, as well as an algebraic interpolation between these two physical limits~\cite{RauSedrakian2023}.

We find that the real part of the $p_1$-mode frequency depends
strongly and non-monotonically on the conversion regime. 
Near $P_t\simeq15\,{\rm MeV/fm^3}$, changing the junction condition from slow to rapid conversion lowers the real part of the $p_1$-mode frequency by approximately $1.4\,{\rm kHz}$ for the same equilibrium star. This $p_1$ frequency shift is more than $15\%$ of the slow-conversion frequency.
By contrast, the imaginary part of the frequency, and hence the damping rate, changes only slightly. At fixed transition pressure, increasing $\Delta\varepsilon$ further enhances the real-frequency separation over the range considered. The principal maximum of this separation occurs in the same transition-pressure region as the strongest local reorganization of the radial and tangential displacements near the interface. This coincidence supports an interpretation in terms of the coupling between the phase boundary and the local $p_1$-mode eigenfunction, although our local diagnostic does not track the frequency separation throughout the full scan. 
Extending the comparison across the EOS grid yields compactness-averaged frequency separations of up to $\sim3.5\,{\rm kHz}$ on disconnected twin-star sequences. These large averaged differences show that the conversion sensitivity remains significant beyond individual stellar configurations, with the strongest effects concentrated in EOSs combining relatively low transition pressures with large density discontinuities.

The $f$-mode behaves differently. 
We find that its complex frequency is only weakly sensitive to the interface condition on a fixed stellar background. We thus calibrate a sub-percent hadronic $f$--$C$ relation using a variety of hadronic EoSs, and map the compactness-averaged departures of both its real-frequency and damping-rate fitting variables across the CSS parameter space.
Departures larger than the hadronic SLy4 residual (i.e.~larger than the sub-percent residual between the hadronic SLy4 $f$--$C$ and the full hadronic fitted function) occur mainly at low $P_t$, large $\Delta\varepsilon$, or both. The deviations are similarly non-monotonic, and are not ordered by either transition parameter alone. Moreover, many departures are comparable to the hadronic SLy4 residual, so their magnitude alone does not provide a clean signature of a phase transition. Changing from slow to rapid conversion modifies the compactness-averaged residual by at most approximately $4\%$ for the real-part fitting variable and $6\%$ for the imaginary-part fitting variable. Thus, the $p_1$-mode primarily probes the local interface response, whereas the $f$-mode primarily probes the global structural departure of the hybrid-star sequence from the hadronic calibration.

The remainder of this paper is organized as follows. Sec.~\ref{sec:setup} describes the EOS construction, polar-perturbation formalism, and interface junction conditions. Sec.~\ref{sec:results} presents the conversion-induced $p_1$-mode frequency separation and the hybrid-star deviations from the hadronic $f$--$C$ relation. Sec.~\ref{sec:discussion} discusses the implications and limitations of the results. 
The detailed numerical procedures are provided in Appendix~\ref{App:numericsprocedures}, and convergence tests can be found in Appendix~\ref{App:numerical_error}.
Unless otherwise stated, we use geometric units with $G=c=1$, Latin letters in indices stand for spacetime coordinates, and bold-faced symbols stand for abstract vectors or matrices.

\section{Stellar models and perturbation setup}
\label{sec:setup}

In this section, we describe the equilibrium stellar models and perturbation formalism used to compute polar quasi-normal modes of hybrid stars with first-order phase transitions. We first introduce the hadronic and hybrid EOS, then summarize the relativistic polar-perturbation equations, and finally discuss the boundary and interface junction conditions. Details of the numerical mode-search procedure are given in Appendix~\ref{App:numericsprocedures}.

\subsection{Equations of state with constant-sound-speed prescription}
\label{Sec:EOS}
We construct equilibrium stellar models from cold, barotropic EOSs $P=P(\varepsilon)$, which relate the pressure $P$ to the energy density $\varepsilon$. In this work, we consider both hadronic stars and hybrid stars that contain a first-order phase transition from a hadronic-matter envelope to a deconfined-quark-matter core.

\begin{figure}
 \centering
 \includegraphics[width=1\linewidth]{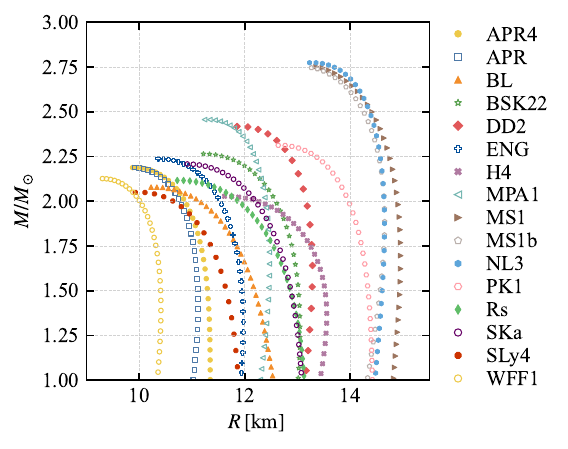}
 \caption{Mass-radius relations for neutron stars with hadronic EOSs. Each curve corresponds to a stellar sequence obtained with one hadronic EOS, as indicated in the legend. Only the stable branch up to the maximum-mass (at least $2\,M_\odot$) is shown, with configurations below $1\,M_\odot$ excluded.}
 \label{fig:hadronic_MR}
\end{figure}

For the hadronic models, we use a collection of tabulated nuclear-matter EOSs\footnote{The tabulated EOSs employed in this work are obtained from \texttt{https://compose.obspm.fr}. For the publications describing the database and interpolation framework, see, e.g., Refs.~\cite{Typel:2013rza, Oertel:2016bki, CompOSECoreTeam:2022ddl}.}, including APR~\cite{Akmal:1998cf}, BL~\cite{Bombaci:2018ksa}, BSK22~\cite{Pearson:2018tkr}, DD2~\cite{Typel:2009sy}, Rs~\cite{Friedrich:1986zza}, SKa~\cite{Kohler:1976fgx}, PK1~\cite{Long:2003dn}, NL3~\cite{Lalazissis:1996rd}, H4~\cite{Grams:2022lci}, and SLy4~\cite{Chabanat:1997un}, together with several piecewise-polytropic EOSs from Ref.~\cite{Read:2008iy}. The corresponding mass--radius relations are shown in~\cref{fig:hadronic_MR}. These hadronic models are used to calibrate the hadronic $f$--$C$ relation discussed in~\cref{sec:hadronic_fc_relation}.
We will otherwise use the SLy4 EOS for a detailed investigation of the impact of first-order phase transition on $p_1$ and $f$-modes.

\begin{figure}
    \centering
    \includegraphics[width=1\linewidth]{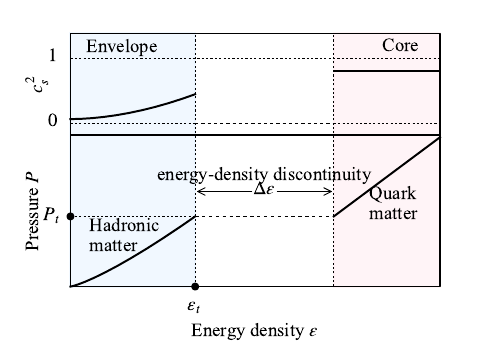}
    \caption{Schematic illustration of a hybrid EOS constructed from a hadronic EOS and a CSS quark-matter EOS. A first-order phase transition is implemented through a Maxwell construction at the transition pressure $P_t$, producing a constant-pressure density jump of magnitude $\Delta\varepsilon$. 
    The blank density interval does not correspond to a finite-thickness layer in the equilibrium star, and no sound-speed value is assigned to it in the upper panel.
    The upper panel shows the corresponding equilibrium sound-speed profile, where the hadronic envelope and quark core are separated by the density-gap region. The shaded blue and red regions indicate the hadronic and quark phases, respectively.}
    \label{fig:cartoon_eos}
\end{figure}

Hybrid EOSs are constructed by matching a tabulated hadronic EOS to a CSS quark-matter EOS through a Maxwell construction~\cite{Alford:2013aca,Alford:2014aya}.
The resulting energy density as a function of pressure is given by
\begin{align}
\varepsilon(P)=
\begin{cases}
\dfrac{P-P_t}{c_{s,\, {\rm con}}^2}
+\varepsilon_{\rm HM}(P_t)
+\Delta \varepsilon,
&
P>P_t,\\
\varepsilon_{\rm HM}(P),
&
P\leqslant P_t,
\end{cases}
\label{eq:CSS EOS}
\end{align}
where $\varepsilon_{\rm HM}(P)$ denotes the hadronic EOS obtained from the tabulated data, $c_{s,\, {\rm con}}$ is the sound speed in the quark phase, $P_t$ is the transition pressure, and $\Delta\varepsilon$ is the energy-density discontinuity across the phase interface~\cite{Alford:2013aca,Alford:2015gna,Han:2018mtj,Christian:2018jyd,Montana:2018bkb,Xie:2020rwg}.
The transition pressure determines the location of the phase boundary inside the star, while the density jump controls the strength of the first-order phase transition. 
Within each phase, the equilibrium sound speed is determined by the local EOS derivative $c_s^2=dP/d\varepsilon$, giving $c_s^2=(d\varepsilon_{\rm HM}/dP)^{-1}$ for $P<P_t$ and $c_s^2=c_{s,\rm con}^2$ for $P>P_t$. At the transition pressure, the energy density jumps between $\varepsilon^-=\varepsilon_{\rm HM}(P_t)$ and $\varepsilon^+=\varepsilon^-+\Delta\varepsilon$. The formal constant-pressure segment connecting these densities in the $P$–$\varepsilon$ diagram has zero slope, $dP/d\varepsilon=0$. However, this segment does not correspond to a finite-thickness layer in an equilibrium star. The two phases meet at a sharp interface, and the sound speeds on its two sides are given by the respective one-sided EOS derivatives. A schematic illustration of this construction is shown in \cref{fig:cartoon_eos}.

Physically, the CSS parametrization provides a simple phenomenological description of deconfined quark matter in which the sound speed is treated as approximately constant over the density range of interest. This parameterization is closely connected to simple phenomenological quark-matter models, such as the MIT bag model~\cite{Chodos:1974pn}. The use of a constant sound speed is also motivated by the expectation that quark matter approaches an approximately conformal EOS at asymptotically high densities, as suggested by perturbative QCD calculations~\cite{Kurkela:2009gj,Kurkela:2010yk}. However, at neutron-star densities, the sound speed is not known from first principles, and we must thus relax the conformal expectation and instead simply require causality and stability: that the sound speed satisfy $0<c_{s,\, {\rm con}}^2\leqslant 1$.

\begin{figure}[tb]
    \centering
    \includegraphics[width=1.0\linewidth]{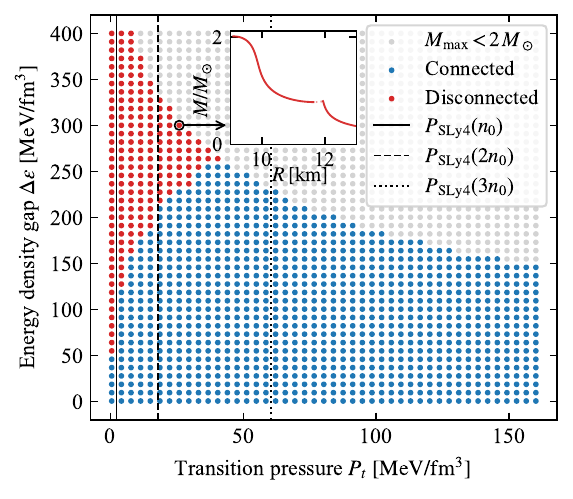}
    \caption{Classification of hybrid-star EOS models in the $(P_t,\Delta\varepsilon)$ parameter space for the SLy4 hadronic baseline. Each point represents one EOS model. Gray points are excluded because they fail to satisfy the observational maximum-mass constraint $M_{\max}\geqslant2M_\odot$. Blue points correspond to EOSs with a single connected stable branch in the mass--radius relation, while red points denote EOSs with two disconnected stable branches, corresponding to the so-called twin-star configurations. The inset shows a representative disconnected mass--radius curve, where the dotted red line is the unstable branch. The solid, dashed, and dotted vertical black lines indicate the SLy4 pressures at nuclear saturation (number) density, $P_{\rm SLy4}(n_0)$, twice nuclear saturation density, $P_{\rm SLy4}(2n_0)$, and three times nuclear saturation density, $P_{\rm SLy4}(3n_0)$, respectively, where we take
    $n_0 \approx 0.159 \ {\rm fm}^{-3}$.
    Models with $P_t<P_{\rm SLy4}(n_0)$, located to the left of the solid vertical line, correspond to sub-saturation transitions and are regarded only as a mathematical stress-test region.
}
    \label{fig:eos_parameter_space_SLY4}
\end{figure}

In the CSS framework, the hybrid EOS is then fully specified by only three parameters, $(c_{s,\, {\rm con}}^2, P_t, \Delta\varepsilon)$. In this work, however, we fix the quark-matter sound speed to the maximally stiff value $c_{s,\, {\rm con}}^2=1$. This choice is not meant to represent the conformal limit or any other limit, but rather to maximize the contrast between the quark core and the hadronic envelope. This choice will thereby make the impact of a first-order phase transition on the oscillation spectrum easier to isolate, at the cost of possibly exaggerating its effect in a real neutron star. 

For the other two free parameters ($P_t$, $\Delta \varepsilon$), we here sample the transition pressure and energy-density discontinuity over $P_t\in[0.5,160]\,{\rm MeV/fm^3}$ and $\Delta\varepsilon\in[0.5,400]\,{\rm MeV/fm^3}$ on the discrete grid shown in \cref{fig:eos_parameter_space_SLY4}, motivated in part by the parameter range considered in Ref.~\cite{Alford:2015gna}.
The portion of the grid with $P_t<P_{\rm SLy4}(n_0)$ is retained only as a mathematical low-transition-pressure stress test of the CSS construction. We do not interpret these sub-saturation models as realistic hadron--quark transitions, and physical conclusions are distinguished from this portion of the parameter space.
We restrict ourselves to the SLy4-based hybrid EOSs satisfying two conditions: (i) the stellar sequence must satisfy the maximum-mass requirement $M_{\rm max}\geqslant 2\,M_\odot$; and (ii) the transition must occur before the maximum-mass configuration, $P_t<P_c(M_{\rm max})$.
In \cref{fig:eos_parameter_space_SLY4}, the grey points are therefore those which do not satisfy the maximum mass condition and are therefore excluded. The retained models are classified as connected or disconnected according to the structure of their stable mass--radius branches: red points denote EOSs with two disconnected stable branches (i.e., twin-star configurations), whereas blue points denote EOSs with a single connected stable branch.

\begin{figure}[tb]
    \centering
    \includegraphics[width=1\linewidth]{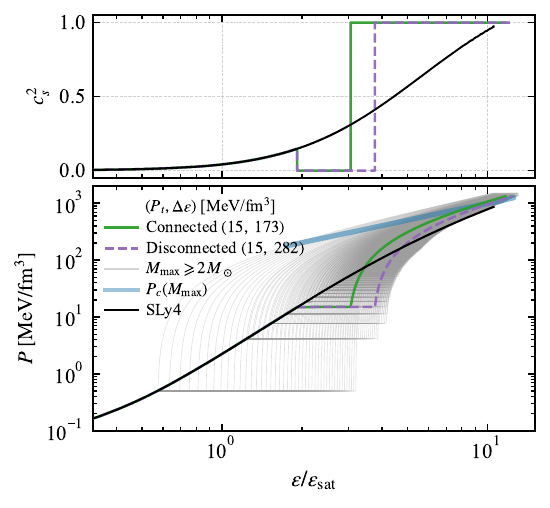}
    \caption{
    Representative hybrid EOSs constructed by hybridizing the SLy4 hadronic EOS. The upper panel shows the squared equilibrium sound speed, $c_{s}^2$, and the lower panel shows the pressure $P$ as functions of the normalized energy density $\varepsilon/\varepsilon_{\rm sat}$, where $\varepsilon_{\rm sat}=\varepsilon_{\rm SLy4}(n_0) \approx 150\ {\rm MeV/fm^{3}}$ is the energy density corresponding to nuclear saturation density $n_0$. The black curve denotes the original, unhybridized SLy4 EOS. The solid green and dashed purple curves show one connected and one disconnected first-order phase-transition EOS with the same transition pressure $P_t$, respectively. The gray curves represent all hybrid EOSs satisfying $M_{\rm max}\geqslant2M_\odot$, corresponding to the colored points in \cref{fig:eos_parameter_space_SLY4}. The blue band indicates the range of central pressures at the maximum-mass configuration, $P_c(M_{\rm max})$, for these allowed hybrid-star sequences.
    }
    \label{fig:representative_EOS_pressure_sound_speed_SLY4}
\end{figure}

\begin{figure}[tb]
    \centering
    \includegraphics[width=1\linewidth]{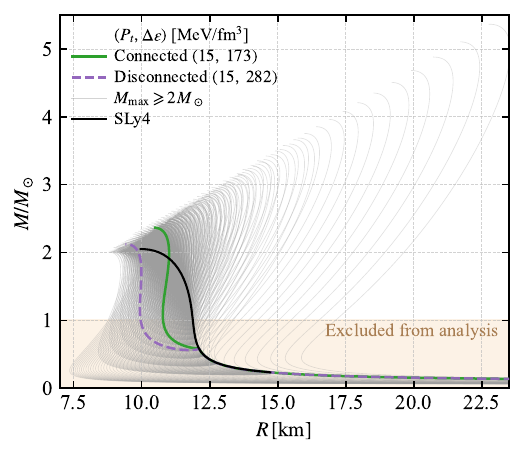}
    \caption{
    Mass–radius relations are shown for the same SLy4-based hybrid EOSs as in \cref{fig:representative_EOS_pressure_sound_speed_SLY4}. The gray curves represent all hybrid EOSs satisfying $M_{\rm max}\geqslant2M_\odot$, while the green solid and purple dashed curves highlight the same representative connected and disconnected EOSs as in \cref{fig:representative_EOS_pressure_sound_speed_SLY4}. The black curve denotes the original SLy4 hadronic sequence. For each EOS, only configurations up to the maximum-mass configuration are shown. Stellar configurations with $M<1\,M_\odot$, indicated by the shaded region, are excluded from the subsequent analysis.
    }
    \label{fig:representative_EOS_MR_SLY4}
\end{figure}

To illustrate the behavior of the EOSs across this parameter space, \cref{fig:representative_EOS_pressure_sound_speed_SLY4} shows the pressure and sound-speed curves for the viable models, with two representative examples highlighted. The blue band indicates the range of central pressures at the maximum-mass configurations, $P_c(M_{\rm max})$, for the retained hybrid-star sequences. The corresponding mass--radius relations are shown in \cref{fig:representative_EOS_MR_SLY4}, where each curve is displayed only up to its maximum-mass configuration. The transition pressure $P_t$ sets where the hybrid sequence branches away from the hadronic one, while the energy-density discontinuity $\Delta\varepsilon$ controls the softening of the sequence and the possible appearance of disconnected branches. These CSS-parameter effects on hybrid-star mass--radius relations are discussed in detail in Refs.~\cite{Dong:2024opz, Alford:2013aca}.

In all analyses below, we consider configurations with a quark-matter core and $M\geqslant1\,M_\odot$ that lie on branches classified as stable according to the conventional turning-point prescription. For EOSs with disconnected sequences, this sample includes the secondary stable branch but excludes the intervening unstable segment\footnote{
Throughout this work, unless explicitly stated otherwise, the terms ``stable'' and ``unstable'' refer only to the conventional turning-point classification of the equilibrium sequence, rather than to conversion-dependent radial dynamical stability. The same turning-point-selected backgrounds are used for the physical slow- and rapid-conversion endpoint conditions and for the intermediate algebraic interpolation. Their radial-stability boundaries are not determined separately.}.
Configurations beyond the final ordinary mass turning point that may remain radially stable specifically in the slow-conversion regime for some CSS parameters are outside the scope of the present survey
\cite{Pereira:2017rmp,Lugones:2021bkm,RauSalaben2023,Mariani:2026xdh}.


\subsection{Polar perturbation equations}
\label{sec:polar_perturbations}

We compute polar quasi-normal modes using the standard Detweiler--Lindblom formulation \cite{Lindblom:1983ps, Detweiler:1985zz} for linear even-parity perturbations of a static, spherically symmetric relativistic star. The background metric $g_{ab}$ is written through the invariant interval $ds^2$ as
\begin{align}
\mathrm{d}s^2
&=g_{ab}\mathrm{d}x^a \mathrm{d}x^b \nonumber \\
&=-e^{\nu(r)}\mathrm{d}t^2
+e^{\lambda(r)}\mathrm{d}r^2
+r^2\mathrm{d}\theta^2
+r^2\sin^2\theta\,\mathrm{d}\phi^2 .
\label{eq:background_metric}
\end{align}
The equilibrium stellar configuration is obtained by solving the Tolman--Oppenheimer--Volkoff (TOV) equations,
\begin{align}
\lambda'
&=
\frac{1-e^\lambda}{r}
+8\pi r e^\lambda \varepsilon ,
\\
\nu'
&=
\frac{e^\lambda-1}{r}
+8\pi r e^\lambda P ,
\\
P'
&=
-\frac{1}{2}(\varepsilon+P)\nu' ,
\end{align}
where a prime denotes differentiation with respect to $r$. We also define the enclosed mass function
\begin{equation}
    m(r)=\frac{r}{2}\left(1-e^{-\lambda}\right),
\label{eq:mass_function}
\end{equation}
so that the total stellar mass is $M=m(R)$, where $R$ is the surface
of the star, which is determined by the condition $P(R)=0$. The additive constant in $\nu$ is fixed by matching the interior metric to the exterior Schwarzschild solution at the stellar surface,
\begin{equation}
    e^{\nu(R)}
    =
    e^{-\lambda(R)}
    =
    1-\frac{2M}{R}.
\label{eq:nu_surface_normalization}
\end{equation}

The perturbed polar metric is decomposed in spherical harmonics and using the harmonic time dependence $e^{+i\omega t}$. 
A stable damped mode can then be written as
\begin{equation}
    \omega
    =
    \omega_{\rm re}
    +
    i \, \omega_{\rm im},
    \qquad
    \omega_{\rm im}
    \equiv
    \operatorname{Im}\omega
    >0\,,
\end{equation}
where throughout $\omega$ denotes the complex angular frequency, which depends on the spherical harmonic index $\ell$ (although we will suppress that here for notational convenience). 
In the Regge--Wheeler gauge~\cite{Regge:1957td, Thorne_Campolattaro_1967, Lindblom:1983ps}, the polar metric perturbation takes the form
\begin{equation}
\delta g_{ab}
=
-r^\ell
\begin{pmatrix}
H_0 e^\nu & i\omega r H_1 & 0 & 0 \\
i\omega r H_1 & H_2 e^\lambda & 0 & 0 \\
0 & 0 & r^2 K & 0 \\
0 & 0 & 0 & r^2\sin^2\theta\,K
\end{pmatrix}
Y_{\ell m}e^{+i\omega t},
\label{eq:rw_gauge_metric_perturbation}
\end{equation}
where $H_0(r)$, $H_1(r)$, $H_2(r)$, and $K(r)$ are radial perturbation functions that also depend on the spherical harmonic index $\ell$ (and that we again suppress here for notational clarity).
In the absence of anisotropic stresses, $H_2=H_0$.

The fluid motion is described by the Lagrangian displacement vector
\begin{equation}
\xi^a
=
r^\ell
\begin{pmatrix}
0 \\
r^{-1}e^{-\lambda/2}W \, Y_{\ell m} \\
-r^{-2}V \, \partial_\theta Y_{\ell m} \\
-r^{-2}\sin^{-2}\theta\,V \, \partial_\phi Y_{\ell m}
\end{pmatrix}
e^{+i\omega t},
\label{eq:fluid_displacement}
\end{equation}
where $W(r)$ and $V(r)$ encode the radial and tangential components of the fluid displacement, respectively, which also depend on the spherical harmonic index $\ell$. One recognizes this decomposition as one in (polar) vector spherical harmonics.

For adiabatic perturbations, the Lagrangian pressure perturbation is related to the Lagrangian perturbation of the baryon number density
$n_b$ by
\begin{equation}
    \Delta P
    =
    \Gamma_1 P\frac{\Delta n_b}{n_b},
\label{eq:lagrangian_pressure}
\end{equation}
where $\Delta=\delta+\mathcal{L}_{\xi}$ denotes the Lagrangian perturbation and $\Gamma_1\equiv (n_b/P)(\partial P/\partial n_b)_{\rm ad}$ is the adiabatic index appropriate to the perturbed fluid, with $\mathcal{L}_{\xi}$ the Lie derivative along the fluid displacement direction.
For the radial perturbation amplitudes used below, $\Delta P$ is encoded in the Detweiler--Lindblom variable $X$ through
\begin{equation}
    \Delta P
    =
    -e^{-\nu/2}r^\ell X .
\label{eq:X_pressure_relation}
\end{equation}

Within each bulk phase, we adopt a barotropic perturbation scheme in which the perturbed fluid follows the same EOS as the corresponding equilibrium background. Consequently,
\begin{equation}
    \Gamma_1=\Gamma_0,
    \qquad
    \Gamma_0
    \equiv
    \frac{\varepsilon+P}{P}
    \frac{\mathrm{d}P}{\mathrm{d}\varepsilon}.
\label{eq:gamma1}
\end{equation}
The derivative in \cref{eq:gamma1} is evaluated separately within each phase. In particular, the derivative is not taken across the discontinuity at the first-order phase interface, which is instead treated through the junction conditions.
In more general perturbation schemes, $\Gamma_1$ is determined by the thermodynamic constraints operating on the oscillation timescale and need not equal the equilibrium adiabatic index $\Gamma_0$ \cite{Thorne1967}.
Composition stratification and finite weak-interaction timescales can then introduce additional buoyancy effects, which are not included here. 
For cold nucleonic stars, a recent relativistic-Cowling calculation found only a few-percent differences in the $f$- and $p_1$-mode frequencies between the beta-equilibrated barotropic and frozen-composition limits \cite{Montefusco:2024xrx}. Recent full-GR work on hyperonic stars has incorporated finite reaction rates through a complex, frequency-dependent dynamical sound speed and found the strongest effects in mode damping and in composition $g$-modes \cite{Ghosh:2026ldg}. Neither result establishes the size of these effects in sharp-interface hybrid matter, which requires a separate calculation.

The Eulerian pressure perturbation is related to the Lagrangian one [\cref{eq:lagrangian_pressure}] through
\begin{equation}
    \delta P=\Delta P-\xi^r P' .
\label{eq:eulerian_lagrangian_pressure}
\end{equation}
Together with the linearized Einstein equations and stress-energy conservation,
\begin{align}
    \delta G_{\mu\nu} &= 8\pi\delta T_{\mu\nu},
    \label{eq:perturbed_einstein}\\
    \delta(\nabla_\mu T^{\mu\nu}) &= 0,
    \label{eq:stress_energy_conservation}
\end{align}
these relations reduce the perturbation problem to a first-order system for the four radial variables
\begin{equation}
    \mathbf{Y}(r)
    =
    \{
    H_1(r),K(r),W(r),X(r)
    \}^{\rm T},
\label{eq:perturbation_vector}
\end{equation}
where the bold-faced symbol stands for an abstract vector of the state variables, while $X=-e^{\nu/2}r^{-\ell}\Delta P$.
The perturbation equations can be written schematically as
\begin{equation}
    \frac{\mathrm{d}\mathbf{Y}}{\mathrm{d}r}
    =
    \mathbf{Q}(r,\omega,\ell)\mathbf{Y},
\label{eq:first_order_system}
\end{equation}
where $\mathbf{Q}$ is a matrix that depends on the background stellar structure, the harmonic index $\ell$, and the complex mode frequency $\omega$. The explicit equations used in our numerical implementation are given in Appendix~\ref{app:interior_equations}; see Refs.~\cite{Lindblom:1983ps,Detweiler:1985zz,Lu:2011zzd,Kruger:2014pva,Gao:2025aqo} for equivalent forms in different conventions
\footnote{The perturbation equations appear in slightly different forms across these references because of differences in conventions and variable definitions. In Ref.~\cite{Lindblom:1983ps}, several typographical errors should be noted: Eq.~(A17) is missing a term $-\ell X/r$, Eq.~(A29) requires the replacement $nr\rightarrow nr(r-2M)$, and Eq.~(B15) was reported by the authors to be seriously misprinted. In Ref.~\cite{Detweiler:1985zz}, the factor $\omega^3$ in Eq.~(6) should instead read $\omega^2$~\cite{Andersson:1995wu}. 
We also note that the coefficient multiplying $V$ in Eq.~(A4) of Ref.~\cite{Gao:2025aqo} should be $1/r^2$. 
We use the corrected expression in our implementation, as shown in Appendix~\ref{app:interior_equations}.
Finally, Ref.~\cite{Lu:2011zzd} adopts a metric convention different from that used in the other references, and they fixed typos in Ref.~\cite{Lindblom:1983ps}.}.

The remaining variables $H_0$ and $V$ are not evolved independently, but are reconstructed algebraically from the evolved variables. Defining $\kappa_{\ell}=\frac{1}{2}(\ell-1)(\ell+2)$, the metric perturbation $H_0$ satisfies
\begin{equation}
\begin{aligned}
&
\left[
(\kappa_{\ell}+1)r
-
\frac{re^{-\lambda}}{2}
(r\lambda'+2)
\right]H_0
\\
&=
r^2e^{-\lambda}
\left[
\omega^2re^{-\nu}
-
\frac{\kappa_{\ell}+1}{2}\nu'
\right]H_1
\\
&\quad
+
\left[
\kappa_{\ell} r
-
\omega^2r^3e^{-\nu}
-
\frac{1}{4}r^2e^{-\lambda}\nu'(r\nu'-2)
\right]K
\\
&\quad
+
4\pi r^2(\varepsilon+P)
\left(
e^{-\lambda/2}\nu'W
+
2\omega^2re^{-\nu}V
\right).
\end{aligned}
\label{eq:H0_constraint}
\end{equation}
The tangential displacement variable $V$ is obtained from
\begin{equation}
\omega^2(\varepsilon+P)V
=
e^{\nu/2}X
+
\frac{1}{r}P'e^{\nu-\lambda/2}W
-
\frac{1}{2}(\varepsilon+P)e^\nu H_0 .
\label{eq:V_algebraic}
\end{equation}
Therefore, the oscillation problem is reduced to an eigenvalue problem for the complex frequency $\omega$, supplemented by the boundary and interface junction conditions discussed below.

\subsection{Boundary conditions at the center and surface}
\label{sec:BC at C and S}

\subsubsection{Regularity at the center}
Since the perturbation equations contain terms proportional to inverse powers of $r$, the center is a regular singular point of the system. The numerical integration is therefore not started exactly at $r=0$.
Instead, we follow the regularity analysis of Detweiler and Lindblom~\cite{Lindblom:1983ps, Detweiler:1985zz}, and construct the nonsingular interior solutions near the center from a power-series expansion.
In particular, for the perturbation variables collected in $\mathbf{Y}(r)$ [\cref{eq:perturbation_vector}], the regular solutions near $r=0$ take the form
\begin{equation}
\mathbf{Y}(r)
=
\mathbf{Y}(0)
+\frac{1}{2}\mathbf{Y}''(0)r^2
+O(r^4).
\label{eq:center_expansion}
\end{equation}
The absence of odd powers follows from the regularity of the scaled Detweiler–Lindblom variables at the origin. Substituting \cref{eq:center_expansion} into the perturbation equations gives algebraic constraints among the central coefficients. At leading order, these constraints relate $H_1(0)$ and $X(0)$ to the two free central amplitudes $W(0)$ and $K(0)$. Thus, there are two linearly independent regular interior solutions.

To determine the quadratic coefficients in \cref{eq:center_expansion}, we next expand the regular background quantities near the center as
\begin{equation}
B(r)=B_c+\frac{1}{2}B_c''r^2+O(r^4),
\qquad
B\in\{\varepsilon,P,\nu\}.
\end{equation}
The absence of terms linear in $r$ follows again from regularity and spherical symmetry, while the coefficients $B_c''$ are determined by the TOV equations and the central EOS. Substituting these background expansions and \cref{eq:center_expansion} into the linear perturbation equations, and collecting the next nonvanishing powers of $r$, gives an algebraic system of the schematic form
\begin{equation}
\mathbf{T}\,\mathbf{Y}(0)''
=
\mathbf{U}\,\mathbf{Y}(0)\,,
\end{equation}
where the matrices $\mathbf T$ and $\mathbf U$ depend on the central background and EOS quantities, as well as on $(\omega,\ell)$. Once the two independent central amplitudes are specified, this system determines the quadratic coefficients required to evaluate the regular solution at the finite starting radius $r_0$.
The explicit expressions for these coefficients are given in Refs.~\cite{Lindblom:1983ps,Detweiler:1985zz,Kruger:2014pva}.

The two-dimensional space of solutions regular at the center can be parametrized by the free amplitudes $W(0)$ and $K(0)$. For each choice of this pair, the leading-order constraints determine $H_1(0)$ and $X(0)$, while the next-order equations determine the quadratic Taylor coefficients $\mathbf{Y}''(0)$. We thus construct two independent basis solutions that are regular at the center by choosing
\begin{equation}
\begin{aligned}
\{W(0),K(0)\}_1&=\{1,\varepsilon_c+P_c\},\\
\{W(0),K(0)\}_2&=\{1,-\varepsilon_c-P_c\}.
\end{aligned}
\end{equation}
The factor $\varepsilon_c+P_c$ provides a convenient relative scaling of the two central amplitudes. The corresponding series solutions are then evaluated at a small but finite radius $r_0$, providing regular initial values for $\{H_1(r_0),K(r_0),W(r_0),X(r_0)\}$ for the outward integrations.

\subsubsection{Surface condition}

At the stellar surface, where the background pressure vanishes, the Lagrangian pressure perturbation must satisfy
\begin{equation}
    \Delta P(R)=0 .
\end{equation}
Using the definition of $X$, this is equivalent to the standard Detweiler–Lindblom surface condition
\begin{equation}
    X(R)=0 .
\label{eq:surface_condition}
\end{equation}
Since \cref{eq:surface_condition} imposes one linear constraint on the four-dimensional perturbation vector $\mathbf{Y}$, the space of surface data that satisfy the boundary condition is three-dimensional. We therefore construct three independent surface basis solutions for the inward integration used in the matching procedure.
The numerical construction of these surface-compatible basis solutions and the subsequent interior--exterior matching are detailed in Appendix \ref{App:interior} and \ref{App:exterior}.

\subsection{Interface junction conditions}
\label{sec:interface_conditions}

Throughout this paper, the phase-transition interface is identified with a hypersurface $\Sigma$ defined by
\begin{equation}
    P(r_t)=P_t ,
\label{eq:phase_equilibrium_condition}
\end{equation}
where $r_t$ denotes the interface radius in the background configuration. For any quantity $A$, we define the jump across
$\Sigma$ as
\begin{equation}
    [A]\equiv A^+-A^- ,
\label{eq:jump_definition}
\end{equation}
where $A^+$ and $A^-$ denote the values evaluated on the exterior and interior sides of the interface, respectively.
When a first-order phase transition occurs inside the star, additional junction conditions are required at the phase boundary to match the perturbation variables across the interface. For convenience, \cref{tab:junction_conditions} summarizes the junction conditions adopted in this work. Their physical interpretation is discussed below.
\begin{table}[b]
\centering
\caption{
Summary of the junction conditions at the first-order phase-transition interface.
The first three conditions impose continuity of the metric variables and normal traction.
The final condition contains the physical slow- and rapid-conversion limits at $\alpha=0$ and $\alpha=1$, respectively, while $0<\alpha<1$ is used as an algebraic interpolation for sensitivity tests.
}
\label{tab:junction_conditions}
\renewcommand{\arraystretch}{1.3}
\begin{tabular}{c c}
\hline\hline
Condition & Interpretation \\
\hline
$[H_1]=0$
&
No-thin-shell polar metric matching
\\
$[K]=0$
&
No-thin-shell polar metric matching
\\
$[\Delta P]=0$
&
Continuity of normal traction
\\
\multirow{3}{*}{
$\displaystyle
\left[
\xi^r-\alpha\frac{\Delta P}{P'}
\right]=0
$
}
&
Slow conversion: $\alpha=0$
\\
&
Algebraic interpolation: $0<\alpha<1$
\\
&
Rapid conversion: $\alpha=1$
\\
\hline\hline
\end{tabular}
\end{table}

In the absence of a surface stress-energy layer, the first and second fundamental forms must be matched across the phase boundary \cite{Israel:1966rt}. In the present Regge--Wheeler gauge [\cref{eq:rw_gauge_metric_perturbation}], the resulting standard no-thin-shell polar metric conditions are
\begin{equation}
    [H_1]=0,
    \qquad
    [K]=0 ,
\label{eq:metric_junction_conditions}
\end{equation}
as used in direct full-GR nonradial calculations with a sharp density discontinuity \cite{Sotani:2001bb,Tonetto:2020bie}.
The variables $H_0$ and $V$ are reconstructed algebraically from the evolved variables through \cref{eq:H0_constraint,eq:V_algebraic}, and they, therefore, do not supply additional independent matching conditions.

Having specified the geometric matching conditions, we now consider the dynamical junction conditions at the perturbed phase boundary.
The conversion timescale depends on microscopic reaction and transport processes and is not determined by the equilibrium EOS alone~\cite{Madsen:1993xx,Alford:2014jha}. We therefore consider the slow- and rapid-conversion limits relative to the oscillation timescale, without assigning a specific conversion rate.
The physical slow- and rapid-conversion conditions for nonradial polar perturbations were derived explicitly in Ref.~\cite{Tonetto:2020bie}; their radial counterparts and the broader radial junction framework are discussed in Refs.~\cite{Pereira:2017rmp,Karlovini:2003xi,RauSedrakian2023}.
A perturbation displaces the interface according to
\begin{equation}
    r_t \rightarrow r_t+\delta r_t ,
\end{equation}
where the relation between the interface displacement $\delta r_t$ and the fluid displacement $\xi^r$ depends on the conversion regime.
Throughout, continuity of the normal traction requires
\begin{equation}
    [\Delta P]=0,
\label{eq:lagrangian_pressure_interface}
\end{equation}
where recall that $\Delta P=\delta P+\xi^r P'$, which is equivalent to the continuity of $X$ in our notation.

In the slow-conversion limit, the conversion timescale is long compared with the oscillation period, and fluid elements initially belonging to one phase remain in that phase throughout the oscillation cycle. 
In other words, the interface behaves as a material surface and moves together with the fluid. 
Since the interface displacement is identical on both sides of the transition, the radial fluid displacement must remain continuous, which requires
\begin{equation}
    [\xi^r]=0 .
\label{eq:slow_conversion_condition}
\end{equation}
The perturbed material interface need not remain at the transition pressure $P_t$, but it moves with the fluid. 

If the phase-conversion timescale is short compared to the oscillation period, the so-called rapid-conversion limit, fluid elements rapidly convert between phases as the star oscillates. Therefore, the interface is no longer advected by the fluid motion as matter near the interface periodically crosses the phase boundary. 
In this case, the interface tracks the instantaneous isobaric equilibrium surface satisfying \cref{eq:phase_equilibrium_condition}.
The perturbed phase-equilibrium condition is
\begin{equation}
    P(r_t+\delta r_t)+\delta P(r_t)=P_t .
\end{equation}
Linearization gives
\begin{equation}
    P'(r_t)\delta r_t+\delta P(r_t)=0 ,
\label{eq:linearized_fast_interface}
\end{equation}
and, using \cref{eq:lagrangian_pressure_interface},
\begin{equation}
    \delta r_t
    =
    \xi^r-\frac{\Delta P}{P'} .
\label{eq:delta_rt_fast}
\end{equation}
Since this perturbed isobaric surface is unique, its displacement must remain continuous across the phase boundary, 
\begin{equation}
    \left[
    \xi^r-\frac{\Delta P}{P'}
    \right]
    =0 .
\label{eq:fast_conversion_condition}
\end{equation}
Here and below, $P'$ inside a jump denotes the one-sided background pressure gradient on the corresponding side of the interface.

For the purpose of testing sensitivity to the interface condition, we also consider the algebraic interpolation appearing in Ref.~\cite{RauSedrakian2023},
\begin{equation}
    \left[
    \xi^r
    -
    \alpha\frac{\Delta P}{P'}
    \right]
    =0,
    \qquad
    0\leqslant\alpha\leqslant1 .
\label{eq:alpha_interpolation}
\end{equation}
The endpoints $\alpha=0$ and $\alpha=1$ recover the physical slow- and rapid-conversion conditions, respectively.
Note that intermediate values $0<\alpha<1$ are only algebraic interpolations in boundary-condition space and are not derived from a finite-rate microscopic conversion model, so they should not be interpreted as calibrated conversion rates.
Moreover, this interpolated condition has not been shown to define proper no-thin-shell junction conditions within the general Karlovini framework \cite{Karlovini:2003xi, RauSedrakian2023}. 
In spite of this, we will still consider this interpolation as a way to understand how the slow- and fast-conversion limits are approached. 

\section{Impact of First-Order Phase Transitions on Polar Oscillations}
\label{sec:results}

A first-order phase transition affects polar oscillations through three parameters varied explicitly in our survey:
(i) the transition pressure $P_t$, which sets the location of the phase interface;
(ii) the density discontinuity $\Delta\varepsilon$, which controls the strength of the transition; and
(iii) the junction condition imposed at the phase interface, which determines how fluid perturbations are matched across the interface.
The purpose of this section is to disentangle the effect of the interface matching prescription at fixed $P_t$ or $\Delta\varepsilon$ and identify which parts of the polar-mode spectrum are most sensitive to each physical channel.
We first study the $p_1$-mode, whose acoustic character makes it particularly sensitive to the interface junction condition, and then turn to the $f$-mode, whose phase-transition imprint is more subtle and is best revealed through deviations from the hadronic $f$--$C$ relation. All numerical results shown below are for quadrupolar polar modes with $\ell=2$.

\subsection{Conversion-induced separation of the $p_1$-mode}
\label{sec:pmode}

In neutron star asteroseismology, the propagation cavities of acoustic and gravity waves are dictated by the Brunt--V{\"a}is{\"a}l{\"a} frequency ($\mathcal{N}$) and the Lamb frequency $L_\ell=[\ell(\ell+1)]^{1/2} c_s/r$, both dependent on the local sound speed, although in different ways~\cite{Unno:1989,Kokkotas:1999bd,Aerts:2010}. 
For $p$-modes, wave propagation is allowed in regions where the squared mode frequency exceeds both\footnote{The familiar inequalities involving $\omega^2$, $L_\ell^2$, and $\mathcal N^2$ should be understood as schematic local/WKB diagnostics, rather than exact global criteria for relativistic stellar eigenmodes.} $\mathcal{N}^2$ and $L_\ell^2$. 
In the piecewise-barotropic bulk phases adopted here, $\Gamma_1=\Gamma_0$, the bulk buoyancy frequency vanishes $\mathcal N^2=0$. The smooth-region acoustic propagation is governed primarily by the sound speed and Lamb frequency, while the discontinuity enters through the interface junction conditions and supports interface-related buoyancy physics.

Previous calculations have associated composition-induced changes in the $p_1$ spectrum with the location of its radial node and with the internal sound-speed profile \cite{Kumar:2023ojk,Sotani:2023zkk}, as discussed in the Introduction. Here, we examine a distinct but related effect: keeping the equilibrium stellar background fixed, we change the junction condition that couples the acoustic perturbations across the sharp phase interface.
The conversion prescription determines how matter responds when displaced radially across the interface, ranging from $[\xi^r]=0$ in the slow-conversion limit to $[\xi^r-\Delta P/P']=0$ in the rapid-conversion limit.
These conditions modify the coupling between the acoustic perturbations on the two sides of the interface, including their relative phases and amplitudes. This effect is particularly important for $p$-modes because, unlike the nodeless $f$-mode, they possess a nontrivial radial nodal structure. The response therefore depends sensitively on the position of the interface relative to the nodes. When the interface lies near a node, even a modest change in the matching condition can shift the node or strongly redistribute the amplitudes of the inner and outer eigenfunction lobes, thereby altering the global phase-matching condition and producing a pronounced frequency shift. Below, we illustrate this interplay using the $p_1$-mode, the lowest-order member of the acoustic-mode family with an internal radial node.

\subsubsection{Dependence on the transition pressure}

\begin{figure}
    \centering
    \includegraphics[width=1\linewidth]{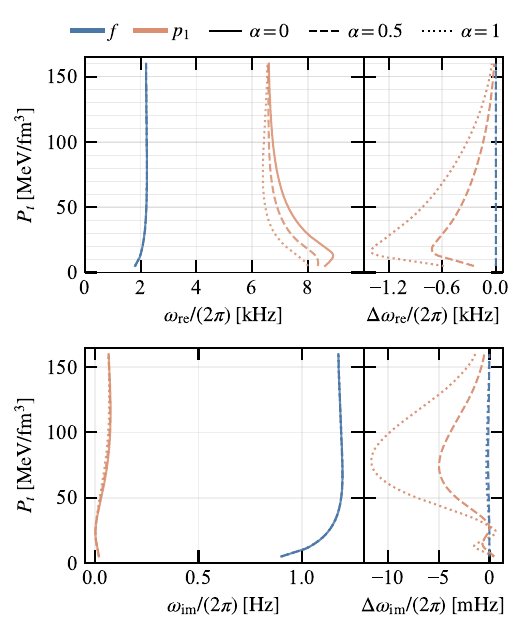}
    \caption{Dependence of the mode frequencies on the transition pressure $P_t$. 
    The left panels show $\omega_{\rm re}/(2\pi)$ and $\omega_{\rm im}/(2\pi)$ for the $f$ (blue) and $p_1$ (orange) modes. Solid, dashed, and dotted curves correspond to $\alpha=0$ (slow), $\alpha=0.5$ (algebraic interpolation), and $\alpha=1$ (rapid), respectively. 
    The right panels show the frequency differences relative to the $\alpha=0$ result at the same $P_t$, using the same linestyles as in the left panels to denote the corresponding values of $\alpha$. Notice the different x-axis scales in the right panels. 
    Results are shown for the SLy4 hybrid-star model with $\Delta \varepsilon =90~\mathrm{MeV/fm^3}$, $P_c=213.66~\mathrm{MeV/fm^3}$, and $c_{s,\, {\rm con}}^2=1$.}
    \label{fig:f_p1_roots_vs_pt}
\end{figure}

To examine how the interface modulates the eigenfunctions and frequencies of $p$-modes, we vary $P_t$ to change where the phase boundary cuts through the fluid eigenfunctions\footnote{In this scan, we fix $P_c$ at a value above the largest transition pressure considered. This representative fixed-$P_c$ choice ensures that all configurations contain a quark core and allows $P_t$ to sample a wide range of core sizes. A lower $P_c$ would truncate the high-$P_t$ part of the scan and could miss the peak of the slow--rapid $p_1$-mode separation.}, while keeping $\Delta\varepsilon$ and $P_c$ fixed. 
\Cref{fig:f_p1_roots_vs_pt} shows the corresponding spectral contrast.
For the $f$-mode, the three interface-condition choices give nearly identical complex frequencies throughout the scan. This suggests that, because its eigenfunction varies smoothly across the star and contains no internal radial node, changing the interface condition produces little redistribution of the $f$-mode displacement between the hadronic envelope and the quark core.
For the $p_1$-mode, however, the real frequency develops a clear conversion-dependent separation that is non-monotonic in $P_t$.
The effect is largest near $P_t\simeq15~{\rm MeV/fm^3}$, with $\Delta\omega_{\rm re}/(2\pi)$ reaching approximately $1.4\,{\rm kHz}$ (more than $15\%$ from the $\alpha=0$ baseline).
The effect becomes weaker toward both the low- and high-$P_t$ limits, while the corresponding differences in the damping rate remain much smaller.
An extension of this scan to lower quark-matter sound speeds, with $\Delta\varepsilon$ and $P_c$ held fixed at the same values as \cref{fig:f_p1_roots_vs_pt}, is presented in Appendix \ref{app:sound_speed}.

\begin{figure}
 \centering
 \includegraphics[width=1.0\linewidth]{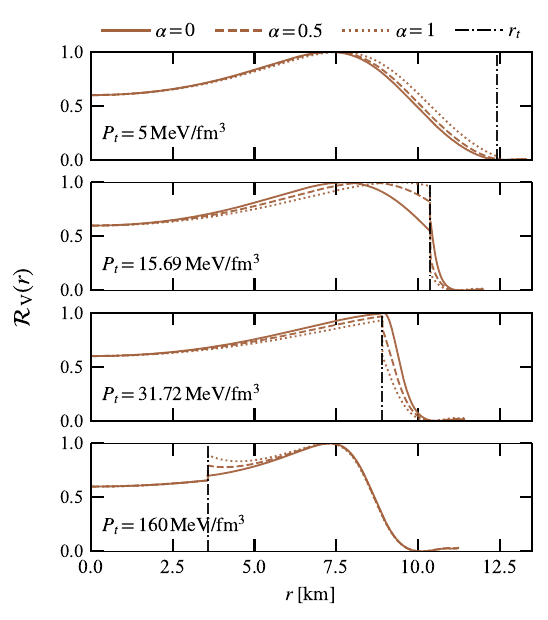}
 \caption{Radial profile of the angular-displacement fraction $\mathcal{R}_{\rm V}(r)$ for the $p_1$-mode eigenfunction. The results are shown for the same hybrid-star sequence as in \cref{fig:f_p1_roots_vs_pt}. From top to bottom, the panels correspond to $P_t=5$, $15.69$, $31.72$, and $160~\mathrm{MeV/fm^3}$. Different line styles denote the interface-condition parameter $\alpha$: solid lines for $\alpha=0$, dashed lines for $\alpha=0.5$, and dotted lines for $\alpha=1$. The black dash-dotted vertical line marks the transition radius $r_t$.}
 \label{fig:RV_profile}
\end{figure}

At fixed $P_t$, all interface matching conditions [\cref{eq:alpha_interpolation}] are applied to the same TOV background because the conversion conditions enter only at the perturbative level.
The resulting mode frequency separation is strongly related to the matching conditions, which are imposed locally at the interface so they act on the values and derivatives of the eigenfunction at that radius.
As the interface moves relative to the radial node, the local displacement structure at the interface can be regarded as a diagnostic of the observed mode-frequency shifts.
To construct a local diagnostic for this comparison, we introduce\footnote{The factor $\ell(\ell+1)$ is introduced because the angular derivatives of the spherical harmonics naturally scale as $\nabla_\perp Y_{\ell m}\sim \sqrt{\ell(\ell+1)}\,Y_{\ell m}/r$, so that $\ell(\ell+1)|V|^2$ provides a more faithful measure of the physical contribution of the tangential fluid motion.},
\begin{equation}
\mathcal{R}_{\rm V}(r)
=
\frac{\ell(\ell+1)|V(r)|^2}
{|W(r)|^2+\ell(\ell+1)|V(r)|^2} \,,
\end{equation}
to measure the fraction of the displacement norm contributed by tangential motion.
Accordingly, $\mathcal{R}_{\rm V}\simeq1$ and 0 indicate tangential-motion and radial-motion dominance, respectively.
For the weakly-damped modes considered here, peaks in $\mathcal R_{\rm V}$ may lie close to a radial-displacement node.
As shown in \cref{fig:RV_profile}, the peak of $\mathcal{R}_{\rm V}$ moves across the phase interface, from its inner side to the outer side, as $P_t$ increases past~$30~{\rm MeV/fm^3}$
\footnote{For $P_t=31.72~{\rm MeV/fm^3}$ in \cref{fig:RV_profile}, the phase interface lies close to the radial nodal crossing.
For $\alpha=0.5$ and $1$, the junction condition permits a discontinuity in $W$ across the interface.
After fixing the arbitrary overall phase of the eigenfunction, the real displacements on either side of the interface have opposite signs and are counted as one interface node.
Because neither one-sided value of $W$ is required to vanish, the corresponding peak of $\mathcal R_{\rm V}$ need not reach unity. 
For more details on node identification, please refer to the Appendix~\ref{App:Mode search}.}.
For most values of $P_t$, the peak locations are nearly independent of $\alpha$, and the overall $\mathcal{R}_{\rm V}$ profiles are also similar. 
A notable exception occurs near $P_t\simeq15~{\rm MeV/fm^3}$, where the peak shifts substantially with the interface condition, from $r\simeq7.5~{\rm km}$ for $\alpha=0$ to $r\simeq10~{\rm km}$ for $\alpha=1$.
This spread indicates distinct internal oscillation structures among the three interface-condition choices and occurs at the same transition pressure as the largest conversion-dependent separation in $\omega_{\rm re}$; see \cref{fig:f_p1_roots_vs_pt}.

\begin{figure}
    \centering
    \includegraphics[width=1\linewidth]{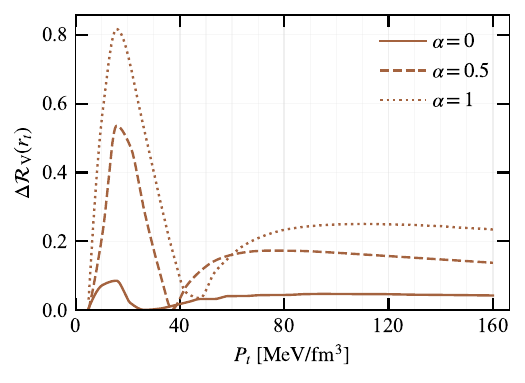}
    \caption{Discontinuity of $\mathcal{R}_{\rm V}$ across the phase interface, shown as a function of the transition pressure $P_t$ for the same hybrid-star sequence as in \cref{fig:f_p1_roots_vs_pt}. The solid, dashed, and dotted curves correspond to $\alpha=0$, $0.5$, and $1$, respectively. The peak near $P_t\simeq15~\mathrm{MeV/fm^3}$ occurs in the same transition-pressure region as the maximum $p_1$-mode separation shown in \cref{fig:f_p1_roots_vs_pt}.}
    \label{fig:RV_jump}
\end{figure}

We quantify the change in the relative radial and tangential contributions across the interface by
\begin{equation}
\Delta \mathcal{R}_{\rm V}(r_t)
=
\left|
\mathcal{R}_{\rm V}(r_t^+)
-
\mathcal{R}_{\rm V}(r_t^-)
\right| \,.
\end{equation}
This jump provides a complementary perspective on this transition. As shown in \cref{fig:RV_jump}, its dependence on $P_t$ exhibits a pronounced peak, signaling the parameter region in which the internal oscillation structure differs most strongly at the interface.
Taken together, \cref{fig:f_p1_roots_vs_pt,fig:RV_profile,fig:RV_jump} show that the largest slow--rapid frequency separation and the largest local displacement mismatch occur in the same transition-pressure region.
Near $P_t\simeq15~{\rm MeV/fm^3}$, the separation between the interface and the nearby peak of $\mathcal R_{\rm V}$ varies most strongly among the three interface-condition choices, with the $\alpha=1$ profile placing the interface closest to the tangentially dominated region. It is also at this transition pressure that \cref{fig:RV_jump} displays the largest $\Delta\mathcal{R}_{\rm V}(r_t)$ and \cref{fig:f_p1_roots_vs_pt} shows the maximal slow--rapid separation in the real part of the $p_1$-mode frequency. The coincidence of these principal maxima suggests that the separation is correlated with how the conversion-dependent matching condition reshapes the eigenfunction locally at the phase boundary. The two quantities do not, however, exhibit a one-to-one relation over the full transition-pressure scan.

Away from the principal peak near $P_t\simeq15~{\rm MeV/fm^3}$, the slow--rapid endpoint frequency separation decreases. The local diagnostic $\Delta\mathcal R_{\rm V}(r_t)$, however, does not follow the endpoint separation monotonically over the full scan.
Around $P_t\simeq30$--$40~{\rm MeV/fm^3}$, each interface condition passes through a nearby minimum in $\Delta\mathcal R_{\rm V}(r_t)$, while a small residual frequency difference remains. These minima occur when the phase interface lies near the peak of the corresponding $\mathcal R_{\rm V}$ profile, with the overlap reached at slightly different transition pressures for the different interface prescriptions. At higher $P_t$, the rapid-conversion value of $\Delta\mathcal R_{\rm V}(r_t)$ rises again and remains appreciable even as the endpoint frequency separation continues to decrease, indicating that a sizable local mismatch need not produce a comparably large global frequency shift.

\subsubsection{Amplification by the density discontinuity}

\begin{figure}
    \centering
    \includegraphics[width=1\linewidth]{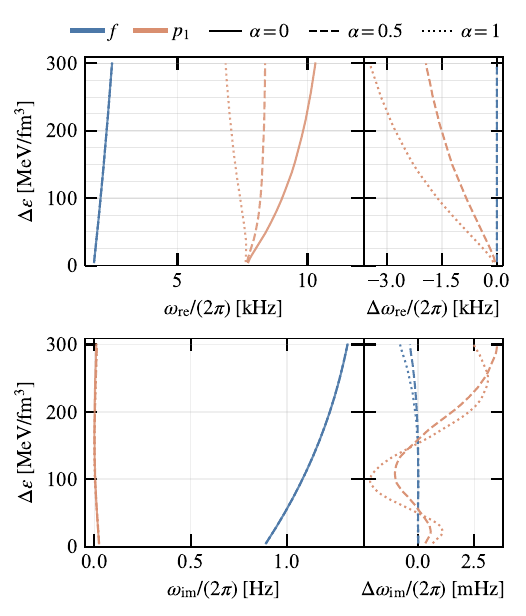}
    \caption{Mode frequencies as functions of the density discontinuity $\Delta\varepsilon$ at $(P_t,\,P_c,\,c_{s,\, {\rm con}}^2)=(15.69~{\rm MeV/fm^3}, 213.66~{\rm MeV/fm^3},1)$. The meanings of all symbols, colors, line styles, and labels are the same as those in \cref{fig:f_p1_roots_vs_pt}.}
    \label{fig:f_p1_roots_vs_dg}
\end{figure}

Having identified $P_t=15.69~{\rm MeV/fm^3}$ as the representative transition pressure at which the largest $p_1$-mode separation occurs in the $P_t$ scan, we now fix $P_t$ and examine how the separation varies across the corresponding family as $\Delta\varepsilon$ is increased. As shown in \cref{fig:f_p1_roots_vs_dg}, the $f$-mode frequencies remain nearly independent of $\alpha$ over the entire range of $\Delta\varepsilon$, indicating that even a stronger discontinuity does not substantially alter its smooth, nodeless displacement. By contrast, the slow--rapid separation of the $p_1$-mode real frequency increases with $\Delta\varepsilon$ in this fixed-$P_t$ sequence. At the representative slice considered here, the different matching prescriptions already produce distinct internal oscillation structures, and increasing the density jump strengthens the resulting frequency separation over the displayed range.

\subsubsection{Summary}

The two representative one-dimensional scans show that both $P_t$ and $\Delta\varepsilon$ affect the conversion-dependent $p_1$-mode separation.
The transition pressure $P_t$ determines where the phase interface lies relative to the radial node of the $p_1$ eigenfunction, and, hence, how strongly the interface conditions differentiate its internal structure. The density discontinuity $\Delta \varepsilon$ controls how strongly this structural difference is reflected in the mode frequency. The $f$-mode, on the other hand, is largely immune to interface conditions, because its smooth global displacement is only weakly modified by the interface matching.

Building on these results, we extend the analysis from representative one-dimensional scans to the surveyed $(P_t,\Delta\varepsilon)$ plane. The results\footnote{During the construction of this map, we found that, in some cases, the number of radial-displacement nodes of a $p$-mode can exceed its nominal radial order. This behavior is discussed in detail in the Appendix~\ref{App:Mode search}.} are summarized in \cref{fig:p1_map}. Numerical sensitivity tests for a representative model are presented in Appendix~\ref{App:numerical_error}. For each EOS, we evaluate the signed compactness-averaged frequency difference,
\begin{align}
    \left\langle\Delta \omega_x/(2\pi)\right\rangle_C
=
\frac{1}{\Delta C}
\int \frac{1}{2\pi}
\left[\omega_x^{\rm rapid}(C)-\omega_x^{\rm slow}(C)\right]\,dC,
\end{align}
where $\omega_x/(2\pi)$, with $x\in\{{\rm re},{\rm im}\}$ denoting the real or imaginary component of the angular frequency. The integral is evaluated over all qualifying compactness intervals corresponding to stable configurations with a quark core and $M\geq 1.0M_\odot$. For connected sequences, $\Delta C=C_{\max}-C_{\min}$. For disconnected sequences, the integral is taken separately over the qualifying stable branches, and $\Delta C$ is defined as the sum of their compactness-interval lengths, excluding the gaps between them.

\begin{figure*}
    \centering
    \includegraphics[width=1\linewidth]{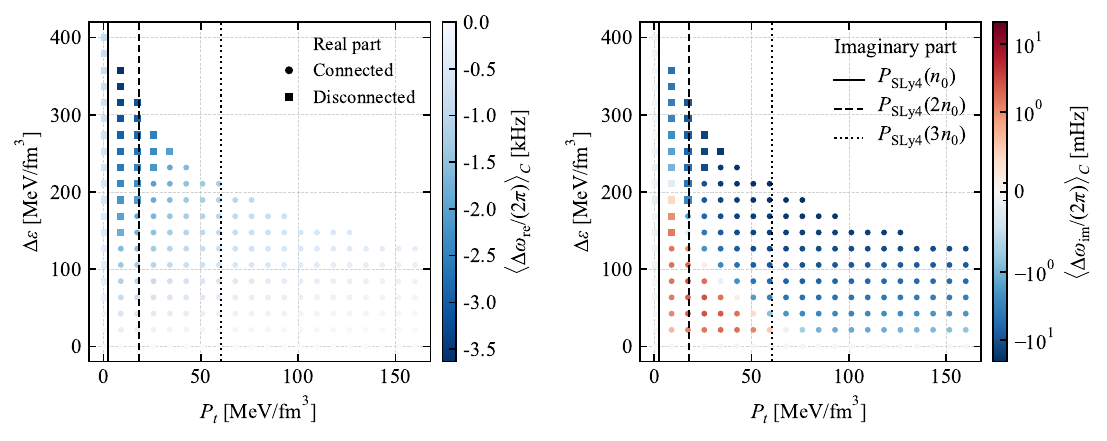}
    \caption{Compactness-averaged differences in the $p_1$-mode frequencies between rapid and slow phase conversion across the SLy4-based hybrid-star EOS parameter space. Each point represents one hybrid EOS in the $(P_t,\Delta\varepsilon)$ plane. Circular and square markers denote EOSs with connected stable branches and disconnected twin-star branches, respectively. The color scales show the signed differences in the real (left, kHz) and imaginary (right, mHz) frequency components. The solid, dashed, and dotted vertical black lines mark the SLy4 pressures at $n_0$, $2n_0$, and $3n_0$, respectively, where $n_0$ is the nuclear saturation density.}
    \label{fig:p1_map}
\end{figure*}

The real-part map shows that rapid conversion generally lowers the $p_1$-mode frequency relative to slow conversion. The strongest reductions occur at relatively low transition pressures and large density discontinuities, with compactness-averaged separations reaching approximately $3.5\,{\rm kHz}$ (30\% of real part of the $p_1$ frequency for slow conversion) on disconnected twin-star branches. At fixed $P_t$, the separation generally increases with $\Delta\varepsilon$, while it is suppressed toward both the smallest transition pressures and the high-$P_t$ portion of the survey. 
The imaginary-part map reaches magnitudes of order $10\,{\rm mHz}$. Positive differences, corresponding to a higher compactness-averaged damping rate for rapid conversion, occur in a wedge-shaped region in the lower-left corner of the map, excluding the boundaries at $P_t=0.5,{\rm MeV/fm^3}$ and $\Delta\varepsilon=0.5,{\rm MeV/fm^3}$. This region extends up to approximately $\Delta\varepsilon=200,{\rm MeV/fm^3}$ and $P_t=60,{\rm MeV/fm^3}$.
Over most of the remaining parameter space, the differences are negative, indicating a lower compactness-averaged damping rate for rapid conversion.

Taken together, these results show that the $p_1$-mode is strongly sensitive to the phase-interface matching conditions. Its conversion-dependent frequency shift relates not only to the transition pressure and density discontinuity, but also to the interplay between the interface location and the internal displacement structure. Therefore, a measurement of the $p_1$ frequency could yield insights into slow- and rapid-conversion in quark cores, although much more work is required to make such a statement quantitative and robust, which is beyond the scope of this paper.

\subsection{Phase-transition imprints on the $f$-mode universal relation}
\label{sec:fC}

In the previous discussion, direct scans in $P_t$, $\Delta\varepsilon$, and $\alpha$ showed that for a fixed stellar background, the $f$-mode frequency is only weakly sensitive to the choice of interface condition (cf.~\cref{fig:f_p1_roots_vs_pt,fig:f_p1_roots_vs_dg}). 
However, the weak sensitivity of the individual $f$-mode frequencies to the interface condition does not preclude systematic sequence-level departures from a tight hadronic universal relation.
Several such relations have been proposed, involving the effective compactness, moment of inertia, and tidal deformability~\cite{Lau:2009bu,Chan:2014kua,Sotani:2021kiw,Zhao:2022tcw}. 
Here, we focus on the compactness-based $f$--$C$ relation, which provides a standard and widely studied benchmark for approximately EOS-insensitive behavior~\cite{Andersson:1997rn,Benhar:2004xg,Tsui:2004qd,Chirenti:2015dda,Wen:2019ouw,Lioutas:2020vzi,Lioutas:2021jbl,Zhao:2022tcw}. 
Full-GR studies have tested standard $f$-mode relations for CSS, Gibbs-mixed, pasta-smoothed, self-bound, and broader exotic-composition samples \cite{Zhao:2022tcw,Zhou:2023nzm,Pradhan:2023zmg,Kumar:2023ojk,Rather:2024mtd}. Approximate universality survives in several of these cases, whereas systematic departures have been found on extended slow-stable branches and in some broader exotic-composition models
\cite{Ranea-Sandoval:2023ixr,Rather:2024mtd}.
Here, we consider a dense two-dimensional $(P_t,\Delta\varepsilon)$ scan constructed from a fixed SLy4 hadronic baseline with $c_{s,\rm con}^2=1$, restricts the sample to the turning-point-selected configurations defined in \cref{fig:eos_parameter_space_SLY4,fig:representative_EOS_MR_SLY4}, adopts the dimensionless fitting variables introduced below, and directly quantifies the difference between the physical slow- and rapid-conversion conditions on identical equilibrium backgrounds.

\subsubsection{Hadronic baseline for the $f$--$C$ relation}
\label{sec:hadronic_fc_relation}

\begin{figure*}
    \includegraphics[width=\textwidth]
    {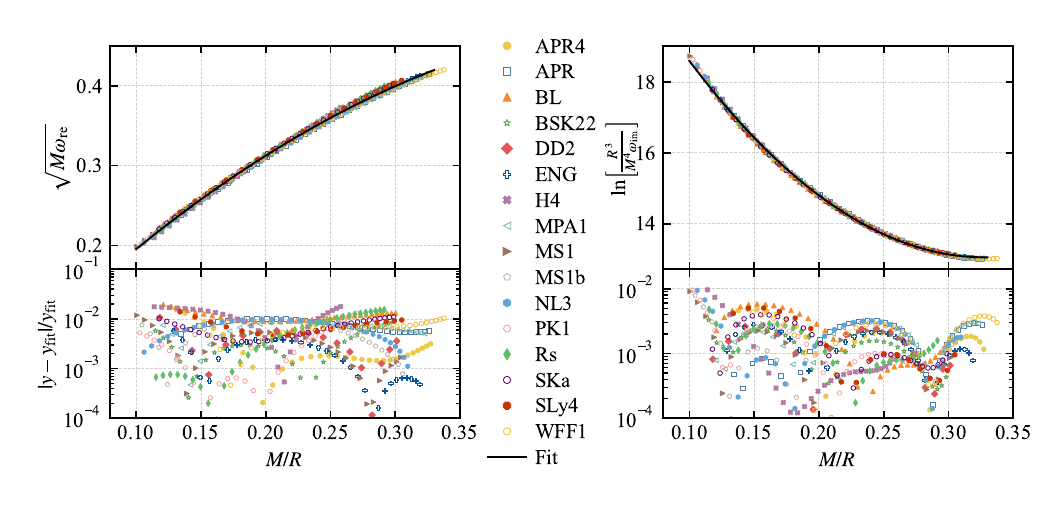}
    \caption{
    Hadronic $f$-mode compactness relations and their fit residuals
    for the EOSs shown in \cref{fig:hadronic_MR}.
    The left and right columns correspond to the fitting variables
    $y_{\rm re}=\sqrt{M\omega_{\rm re}}$ and
    $y_{\rm im}=\ln[R^3/(M^4\omega_{\rm im})]$, respectively.
    The upper panels show the numerical results as functions of the
    compactness $C=M/R$, and the solid black curves show the quadratic
    fits in \cref{eq:hadronic_fitting1}.
    The lower panels show the corresponding pointwise absolute
    normalized residuals of the transformed fitting variables, $|y-y_{\rm fit}|/y_{\rm fit}$, on logarithmic
    scales.
    The fit coefficients and the EOS-averaged,
    compactness-integrated residuals are reported in
    \cref{tab:pure_hadronic_fit}.
    }
    \label{fig:hadronic_fC_relation}
\end{figure*}

To quantify how the sampled hybrid-star sequences depart from a hadronic calibration, we first construct a hadronic baseline.
Relations connecting the $f$-mode frequency and damping time to global stellar properties have been studied extensively since the work of Andersson and Kokkotas~\cite{Andersson:1997rn}. 
Although subsequent analyses have adopted different fitting variables, functional forms, and EOS samples~\cite{Benhar:2004xg,Tsui:2004qd,Lau:2009bu,Chirenti:2015dda,Wen:2019ouw,Lioutas:2020vzi,Lioutas:2021jbl,Zhao:2022tcw}, they consistently find approximately EOS-insensitive behavior when the mode properties are expressed in terms of suitable dimensionless combinations and the stellar compactness.

For the hadronic EOSs shown in \cref{fig:hadronic_MR}, we fit two transformed variables constructed from the real frequency and the positive damping rate of the $f$-mode using the quadratic form
\begin{equation}
y_{\rm fit}
=
a\left(\frac{M}{R}\right)^2
+
b\left(\frac{M}{R}\right)
+
c ,
\label{eq:hadronic_fitting1}
\end{equation}
where the fitted variable is either
\begin{equation}
y_{\rm re}
=
\sqrt{M\omega_{\rm re}},
\qquad {\rm{or}} \qquad
y_{\rm im}
=
\ln\!\left(\frac{R^3}{M^4\omega_{\rm im}}\right).
\label{eq:hadronic_fitting2}
\end{equation}
The best-fit coefficients are summarized in \cref{tab:pure_hadronic_fit}. The last column reports the EOS-averaged, compactness-averaged normalized residual of the transformed fitting variable. We first average the pointwise residual along the stable compactness sequence of each EOS and then average over the hadronic EOS sample,
\begin{equation}
    \left\langle
    \left|
    \frac{y-y_{\rm fit}}{y_{\rm fit}}
    \right|
    \right\rangle_C
    =
    \frac{1}{N_{\rm EOS}}
    \sum_j
    \frac{1}{\Delta C_j}
    \int
    \left|
    \frac{y_j-y_{\rm fit}}
         {y_{\rm fit}}
    \right|
    {\rm d} C .
\label{eq:hadronic_fit_residual}
\end{equation}
where $y$ represents either $y_{\rm re}$ or $y_{\rm im}$.
Here $j$ labels each hadronic EOS listed in \cref{fig:hadronic_MR}, $N_{\rm EOS}$ is the total number of hadronic EOSs included in the fit, and $\Delta C_j \equiv C_{\max,j}-C_{\min,j}$ denotes the compactness interval covered by the stable stellar sequence of the $j$-th EOS.

As shown in \cref{tab:pure_hadronic_fit}, the averaged fractional deviations are below the $1\%$ level for both transformed fitting variables. These values characterize the scatter of the selected $f$--$C$ relations
The numerical-sensitivity tests in Appendix~\ref{App:numerical_error} show that the resolution-dependent changes in the transformed fitting variables are smaller than the dominant structures visible in the deviation maps.
The corresponding normalized residuals of the transformed fitting variables are shown in \cref{fig:hadronic_fC_relation}. 
These hadronic fits will serve as the reference baseline for the following analysis of hybrid-star deviations from the $f$--$C$ relation.

\begin{table}
\caption{
Best-fit coefficients for the hadronic neutron-star models shown in \cref{fig:hadronic_MR}. The last column gives the EOS-averaged, compactness-averaged normalized residual of the corresponding transformed fitting variable.
}
\label{tab:pure_hadronic_fit}
\begin{ruledtabular}
\begin{tabular}{ccccc}
Quantity & $a$ & $b$ & $c$ &
$\langle |\Delta y/y_{\rm fit}| \rangle_C$ \\
\hline
$y_{\rm re}$ & $-1.51126$ & $1.62572$ & $0.0477939$ & $0.574\%$ \\
$y_{\rm im}$ & $106.487$ & $-69.891$ & $24.5191$ & $0.176\%$ \\
\end{tabular}
\end{ruledtabular}
\end{table}

\subsubsection{\label{sec:hybrid_fc_deviation}Hybrid-star deviations from the hadronic baseline}

\begin{figure*}
    \centering
    \includegraphics[width=1\linewidth]{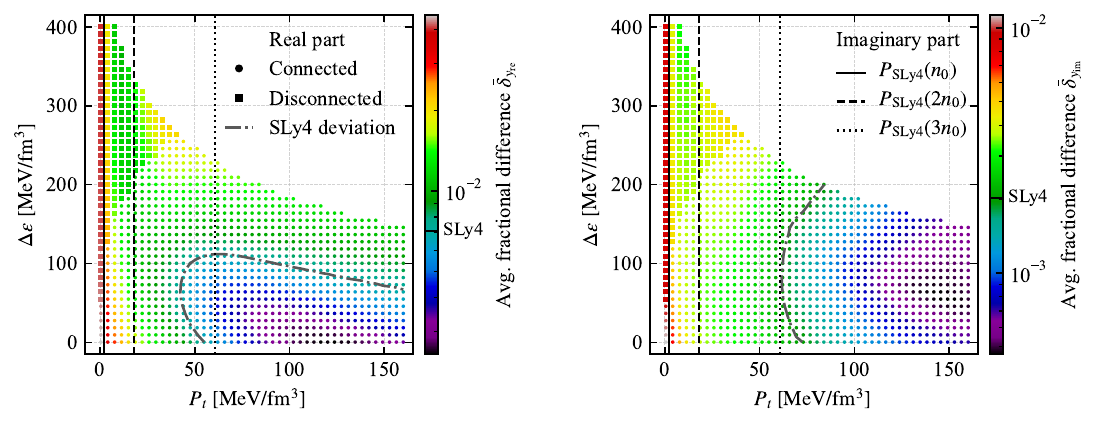}
    \caption{
    Compactness-averaged normalized deviations from the hadronic $f$--$C$ fits across the SLy4-based hybrid-star EOS parameter space. Each point represents one hybrid EOS model in the $(P_t,\Delta\varepsilon)$ plane. Circular markers denote EOSs with connected stable branches, while square markers denote disconnected, twin-star branches. Only stable configurations containing a quark-matter core and satisfying $M\ge1.0\,M_\odot$ are used in the compactness average. The vertical black lines mark the SLy4 pressures at $n_0$, $2n_0$, and $3n_0$. The left panel shows $\bar{\delta}_{y_{\rm re}}$ for the real-part fitting variable $y_{\rm re}$, while the right panel shows $\bar{\delta}_{y_{\rm im}}$ for the imaginary-part fitting variable $y_{\rm im}$. The deviations are computed relative to the quadratic hadronic fits in \cref{eq:hadronic_fitting1,eq:hadronic_fitting2,tab:pure_hadronic_fit} and averaged using \cref{eq:avg_fractional_difference}. The color bar indicates the magnitude of the compactness-averaged fractional deviation. The black tick labeled ``SLy4'' on each color bar marks the corresponding compactness-averaged fractional deviation of the hadronic SLy4 sequence from the same hadronic fit.
    The gray dash-dotted contour in each panel marks $\bar{\delta}_y=\bar{\delta}^{\rm SLy4}_y$; points on the higher-deviation side of the contour have a larger average residual than the hadronic SLy4 sequence.
    }
    \label{fig:avg_fractional_difference_real_imag_SLY4}
\end{figure*}

Using the hadronic $f$--$C$ fits [\cref{eq:hadronic_fitting1,eq:hadronic_fitting2}] as a baseline, we quantify the deviations of hybrid stars across the SLy4-based CSS parameter space. 
For each hybrid EOS model, we use the same compactness-averaged fractional residual as in \cref{eq:hadronic_fit_residual}, but without the outer average over hadronic EOSs,
\begin{equation}
\bar{\delta}_{y}
=
\frac{1}{\Delta C}
\int
\left|
\frac{y-y_{\rm fit}}
     {y_{\rm fit}}
\right|\,{\rm d} C  \,,
\label{eq:avg_fractional_difference}
\end{equation}
to measure the average departure of a given hybrid-star sequence from the hadronic universal relation\footnote{Because the equilibrium background and the available compactness interval vary across the CSS parameter space, the quantity $\bar{\delta}_y$ is defined as a summary over each model's own qualifying hybrid-star branches, but not intended as a pointwise matched comparison at fixed compactness.}.
Recall that here the compactness interval is restricted to stable configurations that contain a quark-matter core and satisfy $M\geqslant1\,M_\odot$. 
If the stable sequence is disconnected, the integral is evaluated over all qualifying compactness intervals, and $\Delta C$ is their combined length.

\begin{figure*}[t]
    \centering
    \includegraphics[width=\textwidth]{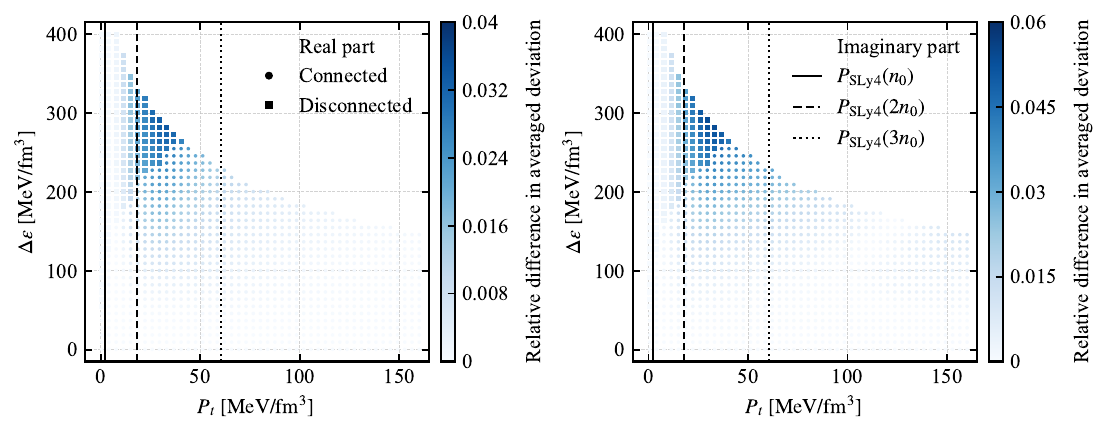}
    \caption{
    Relative rapid--slow difference in the compactness-averaged $f$-mode universal-relation deviations, using the slow-conversion result as the reference value.
    The left and right panels show $d_{y_{\rm re}}$ and $d_{y_{\rm im}}$, respectively, as defined in
    \cref{eq:fmode_relative_endpoint_difference}.
    Each point compares the slow- and rapid-conversion solutions for the same CSS EOS sequence and the same set of turning-point-selected equilibrium configurations.
    Circular and square markers denote connected and disconnected hybrid-star branches.
    The vertical solid, dashed, and dotted lines indicate the SLy4 pressures at $n_0$, $2n_0$, and $3n_0$, respectively.
    }
    \label{fig:fmode_relative_rapid_slow_difference}
\end{figure*}

Since the total deviations obtained with the slow- and rapid-conditions are nearly identical, we use the slow-conversion result as the representative map of the total hybrid-star departure, as shown in \cref{fig:avg_fractional_difference_real_imag_SLY4}. 
The SLy4 markers on the color bars indicate the residual of the full hadronic SLy4 sequence from the same fit and are shown as a common reference scale.
As many of the hybrid-star deviations are comparable to this residual, their magnitude alone may not offer a clean signature of a phase transition. 
The structure of these deviation maps is nevertheless informative. 
The normalized residuals of both transformed fitting variables vary nonmonotonically across the $(P_t,\Delta\varepsilon)$ plane, so the sampled models do not exhibit a simple one-parameter ordering.
Since varying $P_t$ and $\Delta\varepsilon$ changes the interface location, core size, global stellar structure, and qualifying compactness interval together, these maps are interpreted as trends within the SLy4-based CSS family rather than as isolated contributions from individual transition parameters.
The smallest total departure occurs for small $\Delta\varepsilon$ and large $P_t$, where the density discontinuity is weak and the quark core is small. 
The nonmonotonic structure reflects the combined variation of the density jump, the core size and interface location across the CSS survey. 

To quantify the sensitivity of the universal-relation deviations to the interface condition, we define the fractional difference between the compactness-averaged deviations obtained for the same CSS sequence under the slow- and rapid-conversion conditions:
\begin{equation}
    d_y^{\rm UR}(P_t,\Delta\varepsilon)
    =
    \left|
    \frac{
        \bar{\delta}_y^{\rm rapid}
        -
        \bar{\delta}_y^{\rm slow}
    }{
        \bar{\delta}_y^{\rm slow}
    }
    \right| .
\label{eq:fmode_relative_endpoint_difference}
\end{equation}
This statistic measures the relative change in the magnitude of the compactness-averaged transformed-variable residual when the interface condition is changed from slow to rapid conversion.
Since the equilibrium background does not depend on the perturbative conversion condition, the slow and rapid deviations are evaluated over the same stellar configurations and the same qualifying compactness intervals. 
As shown in \cref{fig:fmode_relative_rapid_slow_difference}, the deviations of the real- and imaginary-part transformed variables exhibit their largest rapid--slow differences in approximately the same region, around
$P_t\simeq20$--$40~{\rm MeV/fm^3}$ and $\Delta\varepsilon\simeq220$--$330~{\rm MeV/fm^3}$.
Even there, the relative differences in the compactness-averaged deviation reach only about $4\%$ for $y_{\rm re}$ and $6\%$ for $y_{\rm im}$.
Thus, the choice of physical interface condition has a somewhat minor effect on the $f$-mode universal-relation deviation. 
This is consistent with the preceding mode-frequency comparison in \cref{sec:pmode} and further supports the global character of the $f$-mode and its weak sensitivity to local interface conditions.

\section{Discussion}
\label{sec:discussion}

In this work, we studied polar asteroseismology frequencies of hybrid stars with a sharp first-order phase transition, focusing on how the transition pressure $P_t$, the density discontinuity $\Delta\varepsilon$, and the interface junction condition affect the mode spectrum.
The analysis has been deliberately focused on the SLy4 hadronic baseline to help characterize the effects of the first-order phase transition within a controlled EOS construction.
The main result is that the $p_1$  and $f$-modes probe different aspects of the phase transition. The $p_1$-mode responds directly to the imposed interface condition, whereas the $f$-mode acquires a more global imprint through systematic deviations from the hadronic $f$--$C$ relation.

The behavior of the $p_1$-mode is related to the local coupling between the phase interface and the eigenfunction [\cref{fig:RV_profile}].
Different interface-condition choices can produce sizable frequency separation for the same background stellar model, reaching a slow--rapid real-frequency separation of approximately $1.4\,{\rm kHz}$ for the $p_1$-mode in the representative transition-pressure scan at fixed density gap. At fixed $P_t$, increasing $\Delta\varepsilon$ further enhances the separation over the displayed density-jump range.
The size of this separation is jointly affected by $P_t$ and $\Delta\varepsilon$ (\cref{fig:f_p1_roots_vs_pt,fig:f_p1_roots_vs_dg}).
The principal maximum of the endpoint separation occurs in the same transition-pressure region as the principal maximum of the local displacement-mismatch diagnostic. This qualitative co-variation is consistent with a role for local eigenfunction reorganization, although the local diagnostic does not track the endpoint separation over the full scan.

The two-dimensional survey in \cref{fig:p1_map} identifies a preferred region of strong conversion sensitivity of $p_1$-mode at relatively low transition pressures and large density discontinuities, with compactness-averaged frequency separations reaching approximately $3.5\,{\rm kHz}$ on disconnected twin-star branches. The averaged frequency separation is suppressed at both extremes of the transition-pressure range and generally increases with the density discontinuity at fixed $P_t$. Sensitivity to the interface conversion prescription is widespread across the surveyed parameter space. Regions with larger $p_1$-mode frequency separations may offer better prospects for observationally distinguishing slow and rapid conversion.

The $f$-mode shows a different kind of sensitivity, appearing through small but systematic deviations from the hadronic universal relation.
Relative to the deviation of the hadronic SLy4 sequence, larger departures occur mainly in the low-$P_t$ and/or large-$\Delta\varepsilon$ portions of the sampled parameter space, whereas moderate-to-high $P_t$ combined with a sufficiently small density discontinuity generally gives comparable or smaller deviations. 
The rapid--slow differences in the averaged deviations remain at the few-percent level.
As the departures from the hadronic baseline are not organized by a single parameter and are nearly consistent in slow- and rapid-conversion, the $f$-mode does not provide the same local probe of interface dynamics as the $p_1$-mode. 
Instead, it can act as a consistency test for whether a sequence follows the hadronic universal-relation baseline.

The observational implications of our results require further study. A conversion-sensitive $p_1$-mode frequency could constrain the interface conversion regime only if the relevant pressure mode is sufficiently excited, sufficiently long-lived, and resolvable in the GW spectrum
\cite{Breschi:2022ens,Miao:2017qot}. 
However, no confirmed gravitational-wave detection of a neutron-star $p$-mode has yet been reported.
Interpreting such a frequency in terms of conversion physics would additionally require controlling degeneracies with the hadronic EOS and the stellar parameters. 
For discontinuity or interface modes, previous work has quantified resonant tidal excitation and possible inspiral signatures \cite{Lau:2020bfq,Miao:2023jqe,Counsell:2025hcv,
Pereira:2025xsi}. The analogous excitation, damping, and
resolvability problem for the conversion-sensitive $p_1$-mode
considered here remains open.

These questions are particularly important for postmerger remnants, where several fluid modes may be present. Numerical-relativity studies have shown that a hadron--quark phase transition can alter dominant and subdominant postmerger spectral features, produce delayed frequency changes, and modify the remnant dynamics \cite{Most:2018eaw,Bauswein:2018bma,Weih:2019xvw,Espino:2023llj}. 
The finite-temperature location and shape of the phase boundary can itself substantially affect the postmerger spectrum \cite{Blacker:2023afl}. These hot, rotating-remnant features should not be identified directly with the cold, nonrotating $f$- and $p_1$-mode modes calculated here. The $f$-mode departures from the hadronic universal relation are likely to be subtler: rather than providing a standalone signature in a single event, they may be more useful as part of a broader asteroseismology consistency test across multiple sources. 
Quantifying these possibilities will require waveform modeling, mode-excitation estimates, damping-time calculations, and detector-sensitivity studies in the future.

\begin{acknowledgments}
We thank Hajime Sotani for helpful comments that improved the clarity and precision of the manuscript. Z. Z. Dong would like to thank Shu Yan Lau for helpful suggestions. We acknowledge support from the Simons Foundation through Award No.~896696, the Simons Foundation International through Award No.~SFI-MPS-BH-00012593-01, and the NSF through Grant No.~PHY-25-12423.
\end{acknowledgments}

\appendix

\section{Dependence on the quark-matter sound speed}
\label{app:sound_speed}

\begin{figure}
    \centering
    \includegraphics[width=1\linewidth]{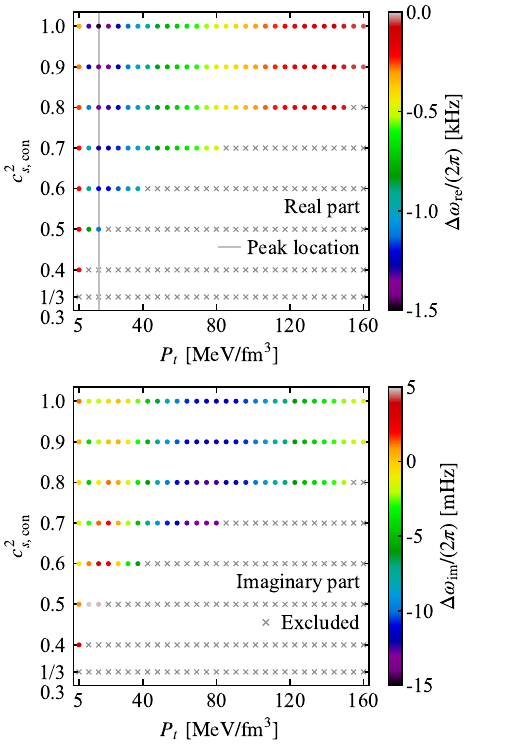}
    \caption{Dependence of the slow--rapid $p_1$-mode frequency separation on the transition pressure and quark-matter sound speed for the SLy4-based hybrid-EOS family with $\Delta\varepsilon=90\ \mathrm{MeV/fm^3}$ and $P_c=213.66\ \mathrm{MeV/fm^3}$. Colored circles show $\Delta\omega_{\mathrm{re}}/(2\pi)$ in kHz (top) and $\Delta\omega_{\mathrm{im}}/(2\pi)$ in mHz (bottom). Gray crosses indicate EOSs that fail the equilibrium-sequence selection criteria. The solid gray line marks $P_t\simeq15\ \mathrm{MeV/fm^3}$, the sampled location of the largest real-part separation for most series with $0.5\leqslant c_{s,\mathrm{con}}^2\leqslant1$.}
    \label{fig:sound_speed_dependence}
\end{figure}

This appendix complements the analysis in \cref{fig:f_p1_roots_vs_pt} by examining how the slow/rapid $p_1$-mode frequency separation depends on the quark-matter sound speed.
We use the SLy4 hadronic EOS and the constant-sound-speed parametrization of the quark phase, with $\Delta\varepsilon=90\ \mathrm{MeV/fm^3}$ and $P_c=213.66\ \mathrm{MeV/fm^3}$.
The transition pressure $P_t$ is sampled between $5$ and $160\ \mathrm{MeV/fm^3}$.
We vary $c_{s,\mathrm{con}}^2$ and calculate the $p_1$-mode frequencies in both the slow- and rapid-conversion limits ($\alpha=0$ and $1$, respectively). We apply the same equilibrium-sequence selection criteria as in the main text. We require that the EOS support a maximum mass $M_{\max}\geqslant2M_\odot$, and the configuration at the chosen $P_c$ contain a quark core, have $M\geqslant1M_\odot$, and satisfy the conventional stable branch requirement.

\Cref{fig:sound_speed_dependence} shows the signed frequency differences,
\begin{equation}
    \Delta\omega_{\mathrm{re/im}}
    =
    \omega_{\mathrm{re/im}}^{\mathrm{rapid}}
    -
    \omega_{\mathrm{re/im}}^{\mathrm{slow}},
\end{equation}
with $\Delta\omega_{\mathrm{re}}/(2\pi)$ expressed in kHz and $\Delta\omega_{\mathrm{im}}/(2\pi)$ in mHz.
The real-part difference is negative throughout the sample, indicating a lower $p_1$-mode frequency for rapid conversion.
Here, the peak separation refers to the largest sampled $|\Delta\omega_{\mathrm{re}}|/(2\pi)$ as $P_t$ varies at fixed $P_c$ and $c_{s,\mathrm{con}}^2$.

Most series peak at $P_t\simeq15\ \mathrm{MeV/fm^3}$, marked by the solid gray line. 
Although finer sampling in $P_t$ may reveal some shifts in the precise peak locations, the present results indicate that the maxima remain broadly concentrated in this vicinity.
The peak separation decreases from approximately $1.47$~kHz at $c_{s,\mathrm{con}}^2=1$ to $1.11$~kHz at $c_{s,\mathrm{con}}^2=0.5$.
The imaginary-part difference changes sign across the parameter region and reaches magnitudes of order $10$~mHz. 
At lower sound speeds, however, fewer EOSs satisfy the selection criteria, restricting the available range of $P_t$ and potentially leaving part of this dependence unresolved.

These results support the main-text conclusion that the $p_1$-mode frequency is strongly sensitive to the phase-interface matching conditions.
A substantial slow--rapid separation persists at lower quark-matter sound speeds in the admissible models examined here, although its magnitude depends on the stiffness of the quark EOS.

\section{Numerical sensitivity tests}
\label{App:numerical_error}
\begin{figure}
    \centering
    \includegraphics[width=1\linewidth]{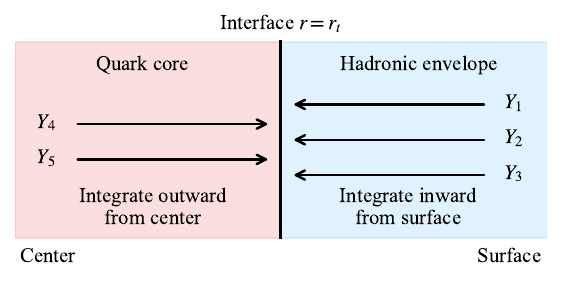}
    \caption{
    Schematic illustration of the numerical matching procedure for a
    hybrid star.
    The three surface-compatible basis solutions
    $\mathbf{Y}_1$, $\mathbf{Y}_2$, and $\mathbf{Y}_3$ are initialized
    near the stellar surface and integrated inward through the
    hadronic envelope.
    The two center-regular solutions $\mathbf{Y}_4$ and
    $\mathbf{Y}_5$ are integrated outward through the quark core.
    The two solution families are matched at the phase-transition
    interface $r=r_t$ using the interface junction conditions.
    }
    \label{fig:numerical_matching_scheme}
\end{figure}

\begin{figure}
    \centering
    \includegraphics[width=\linewidth]
    {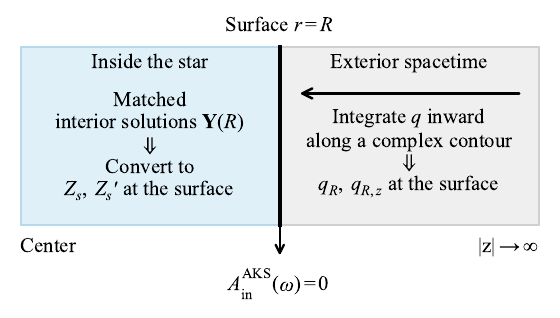}
    \caption{
    Schematic illustration of the exterior matching procedure.
    The interior solution provides the Zerilli surface data $(Z_s,Z_s')$, while the exterior phase-amplitude integration yields $(q_R,q_{R,z})$.
    A quasi-normal mode is obtained by imposing $A_{\rm in}^{\rm AKS}=0$.
    }
    \label{fig:surface_mode_condition}
\end{figure}

We begin by briefly summarizing the computational workflow needed to contextualize the numerical sensitivity tests presented below. For a given EOS and central pressure, we first solve the TOV equations to construct the equilibrium stellar background. For each trial complex frequency $\omega$, the polar perturbation equations are integrated from both the center and the surface, and the two solution families are matched at the phase-transition interface using the junction conditions described in \cref{sec:interface_conditions}, as illustrated schematically in \cref{fig:numerical_matching_scheme}. The resulting interior solution is then converted to Zerilli data at the stellar surface and matched to an exterior solution satisfying the purely outgoing-wave condition at infinity, as summarized in \cref{fig:surface_mode_condition}. The quasi-normal-mode frequencies are obtained by searching for the complex frequencies that satisfy this outgoing-wave condition. The detailed numerical implementation, including the interior basis construction, surface treatment, exterior phase-amplitude integration, and mode-search procedure, is given in Appendix \ref{App:numericsprocedures}.

To assess the numerical accuracy of the calculation, we examine two distinct sources of numerical sensitivity:
(i) the replacement of the formal zero-pressure surface by a finite-pressure cutoff, and
(ii) the discretization of the TOV background, interior perturbation, and exterior AKS integrations.
We perform these tests for a representative SLy4-based hybrid-star configuration selected from the region of \cref{fig:p1_map} exhibiting the largest slow–rapid $p_1$-mode frequency separation. The EOS parameters are $P_t=8.89\ \mathrm{MeV/fm^3}$, $\Delta\varepsilon=357.95\ \mathrm{MeV/fm^3}$, and $c_{s,\mathrm{con}}^2=1$. We choose $P_c=242.72\ \mathrm{MeV/fm^3}$, corresponding to the largest separation, approximately $4.5\ \mathrm{kHz}$, among the central pressures sampled for this EOS. Both the slow ($\alpha=0$) and rapid ($\alpha=1$) conversion limits are examined using the same equilibrium background.
The production calculations in the main text use
\begin{align}
    \frac{P_{\rm cut}}{P_c}&=10^{-14},\\
    \qquad
    h_{\rm TOV}&=1.0\times10^{-3}~{\rm km},\\
    \qquad
    h_{\rm in}&=1.0\times10^{-4}~{\rm km},\\
    \qquad
    h_{\rm ex}&=4.0\times10^{-1},
\label{eq:production_numerical_settings}
\end{align}
where $h_{\rm TOV}$ and $h_{\rm in}$ are the radial step sizes used for
the TOV and interior-perturbation integrations, respectively, and
$h_{\rm ex}=|\Delta z|$ is the dimensionless step size along the exterior complex contour.

For any numerical quantity $\gamma$, we define its fractional
difference from a reference value as
\begin{equation}
    \delta_\gamma
    \equiv
    \frac{\gamma-\gamma_{\rm ref}}
         {\gamma_{\rm ref}}.
\label{eq:numerical_fractional_difference}
\end{equation}
For the surface-cutoff test, $\gamma_{\rm ref}$ denotes the result
obtained using the production cutoff.
For each step-size scan, it denotes the result obtained at the finest
sampled resolution of the component being varied.

\subsection{Finite-pressure surface cutoff}
To test the sensitivity to the numerical surface cutoff, we increased the cutoff over ${P_{\rm cut}}/{P_c} = 10^{-13},\,10^{-12},\,10^{-11}$, and $10^{-10}$,
while keeping all integration resolutions fixed at their production values.
All differences in this scan are evaluated relative to the production choice $P_{\rm cut}/P_c=10^{-14}$.

\begin{figure}
    \centering
    \includegraphics[width=1\linewidth]
    {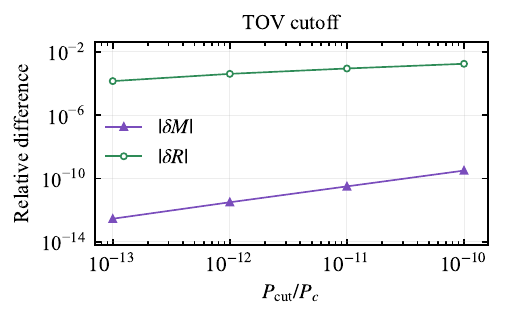}
    \caption{Sensitivity of the stellar mass and radius to the finite-pressure cutoff used to define the numerical stellar surface. Purple triangles and green open circles show the absolute fractional differences in the stellar mass and radius, respectively. The TOV background is identical for the slow- and rapid-conversion limits. All differences are evaluated relative to the production choice $P_{\rm cut}/P_c=10^{-14}$, which is omitted because its differences vanish by definition. All other numerical parameters are held at their production values.
    }
    \label{fig:surface_cutoff_sensitivity_TOV}
\end{figure}

\begin{figure}
    \centering
    \includegraphics[width=1\linewidth]{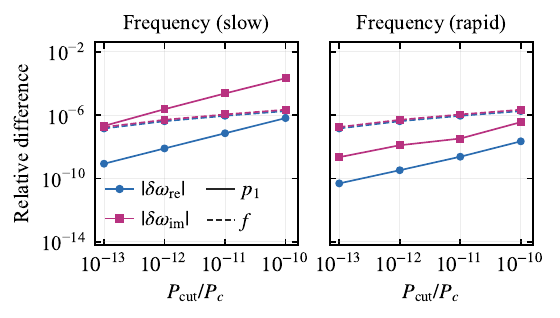}
    \caption{Sensitivity of the complex mode frequencies to the finite-pressure cutoff used to define the numerical stellar surface. The left and right panels show the slow- and rapid-conversion limits, respectively. Blue circles and magenta squares denote the absolute fractional differences in $\omega_{\rm re}$ and $\omega_{\rm im}$, respectively, for the $p_1$-mode (solid curves) and the $f$-mode (dashed curves). The reference cutoff and fixed numerical settings are the same as \cref{fig:surface_cutoff_sensitivity_TOV}.}
    \label{fig:surface_cutoff_sensitivity_frequency}
\end{figure}

As shown in \cref{fig:surface_cutoff_sensitivity_TOV}, the stellar mass is essentially insensitive to the cutoff over the tested range, with $|\delta_M|\lesssim10^{-9}$ even for $P_{\rm cut}/P_c=10^{-10}$.
The radius is more directly affected by moving the numerical surface, but its fractional change remains below approximately $2\times10^{-3}$ at the coarsest cutoff.
The mode frequencies are substantially less sensitive than the radius. 
In \cref{fig:surface_cutoff_sensitivity_frequency}, the relative errors in $\omega_{\rm re}$ and $\omega_{\rm im}$ remain below a few $10^{-4}$ for the $p_1$-mode in both conversion limits.
For the $f$-mode, the corresponding errors remain below a few $10^{-6}$.
Thus, even increasing the cutoff by as many as four orders of magnitude produces only small changes in the quantities studied in the present analysis, which are not large enough to affect the conclusions of this work.

\subsection{Step-size sensitivity}
\begin{figure}
    \centering
    \includegraphics[width=1\linewidth]
    {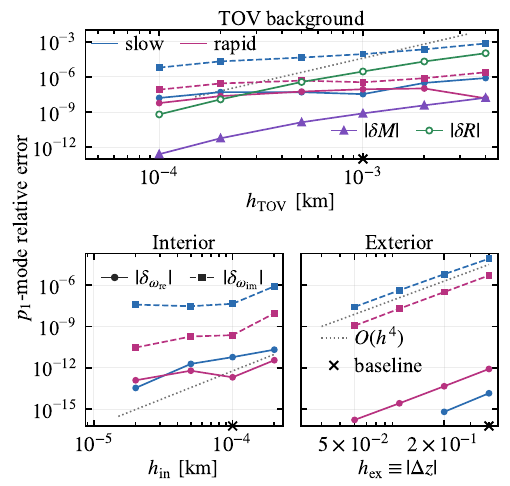}
    \caption{Resolution-sensitivity test for the $p_1$-mode. The TOV panel shows $|\delta_M|$, $|\delta_R|$, $|\delta_{\omega_{\rm re}}|$, and $|\delta_{\omega_{\rm im}}|$. The interior and exterior panels show the two frequency errors. Blue and magenta frequency curves denote slow and rapid conversion. Solid curves with circles and dashed curves with squares represent the real- and imaginary-part errors, respectively. Purple filled triangles and green open circles denote the mass and radius errors, shared by both conversion limits. Black crosses mark the production resolutions. Gray dotted lines show representative $O(h^4)$ slopes only as visual guides and do not represent fitted convergence orders. The exterior resolution is the dimensionless contour step $h_{\rm ex}=|\Delta z|$.
    }
    \label{fig:numerical_convergence_pmode}
\end{figure}

\begin{figure}
    \centering
    \includegraphics[width=1\linewidth]
    {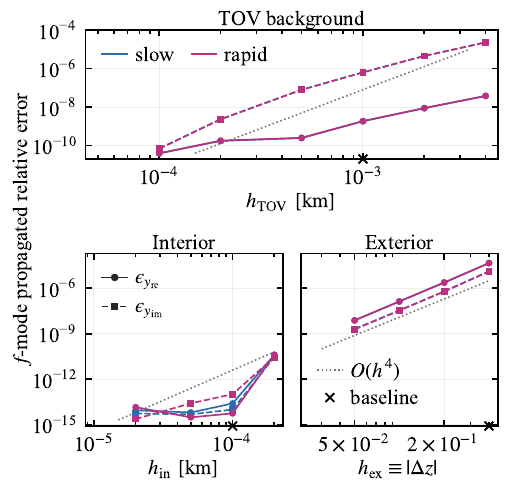}
    \caption{Resolution-sensitivity test for the transformed $f$-mode fitting variables. The three panels show the propagated relative errors $\epsilon_{y_{\rm re}}$ and $\epsilon_{y_{\rm im}}$ obtained by varying the TOV, interior, and exterior resolutions, respectively. Blue and magenta curves denote slow and rapid conversion. Solid curves with circles show $\epsilon_{y_{\rm re}}$, while dashed curves with squares show $\epsilon_{y_{\rm im}}$. The color, line, and marker conventions, the $O(h^4)$ guides, and the definition of the exterior resolution are the same as \cref{fig:numerical_convergence_pmode}.
    }
    \label{fig:numerical_error_budget_fmode}
\end{figure}

We next vary independently the numerical resolutions of
(i) the TOV background construction,
(ii) the interior perturbation integration, and
(iii) the exterior AKS phase integration.
All three components are integrated using fixed-step fourth-order Runge--Kutta (RK4) schemes.
In each scan, only one numerical step size is varied over the ranges shown in \cref{fig:numerical_convergence_pmode} ($p_1$-mode) and \cref{fig:numerical_error_budget_fmode} ($f$-mode), while the other components are held fixed at high-resolution values.
The production values are marked by crosses on the horizontal axes.
The dotted $O(h^4)$ segments in the figures are included only as visual guides to the nominal RK4 scaling, and no formal observed convergence order is inferred from them, as this is not necessary here.

For the $p_1$-mode, the TOV-background scan is characterized by $|\delta_M|$ and $|\delta_R|$, together with the induced frequency differences
$|\delta_{\omega_{\rm re}}|$ and
$|\delta_{\omega_{\rm im}}|$.
The interior and exterior scans are characterized directly by the frequency differences, as shown in \cref{fig:numerical_convergence_pmode}.
For the $f$-mode, the quantities entering the universal-relation analysis are the transformed fitting variables $y_{\rm re}$ and $y_{\rm im}$ defined in \cref{eq:hadronic_fitting2}, rather than the raw frequencies.
We therefore monitor
\begin{equation}
    \epsilon_{y_{\rm re}}
    \equiv
    \left|
    \frac{
    y_{\rm re}(h)-y_{{\rm re},{\rm ref}}
    }{
    y_{{\rm re},{\rm ref}}
    }
    \right|,
    \qquad
    \epsilon_{y_{\rm im}}
    \equiv
    \left|
    \frac{
    y_{\rm im}(h)-y_{{\rm im},{\rm ref}}
    }{
    y_{{\rm im},{\rm ref}}
    }
    \right|.
\label{eq:fmode_transformed_numerical_differences}
\end{equation}
To leading order, these variations can be propagated from the numerical differences in $M$, $R$, and $\omega$ as
\begin{equation}
    \epsilon_{y_{\rm re}}
    \simeq
    \frac{1}{2}
    \left|
    \delta_M+\delta_{\omega_{\rm re}}
    \right|,
\label{eq:yreal_numerical_propagation}
\end{equation}
and
\begin{equation}
    \epsilon_{y_{\rm im}}
    \simeq
    \frac{1}{|y_{{\rm im},{\rm ref}}|}
    \left|
    3\delta_R
    -4\delta_M
    -\delta_{\omega_{\rm im}}
    \right|.
\label{eq:yimag_numerical_propagation}
\end{equation}
The resulting resolution sensitivities are shown in \cref{fig:numerical_error_budget_fmode}.

The scans show an overall decrease in resolution sensitivity as the step sizes are reduced, although the curves are not strictly monotonic at every sampled point and a single fourth-order power law does not describe the full range.
This behavior is expected because the measured differences contain not only the local RK4 truncation error, but also contributions from the surface interpolation, the numerical localization of the sharp phase interface, the interior--exterior matching, and the complex-frequency root search.

For the TOV scan, the changes in $M$ and $R$ remain small throughout the production-and-finer range.
The induced relative fractional differences in the real and imaginary parts of the $p_1$-mode frequency remain at or below a few $10^{-4}$.
For the transformed $f$-mode variables, the corresponding propagated fractional differences remain below approximately $5\times10^{-6}$ near and below the production value of $h_{\rm TOV}$.

The interior scan shows a similar reduction toward finer resolution.
The $p_1$-mode relative frequency errors remain below approximately $10^{-7}$, while the measured propagated differences of the $f$-mode fitting variables are of order $10^{-13}$ or smaller.

The exterior AKS scan exhibits a smooth reduction for quantities that remain above the numerical floor.
For the $p_1$-mode, the relative errors in $\omega_{\rm re}$ and $\omega_{\rm im}$ remain below approximately $10^{-12}$ and $10^{-4}$, respectively, over the sampled range in both conversion limits.
Some real-frequency differences fall below $10^{-15}$, where finite-precision and root-search effects become relevant.
All nonzero values are included in the figure; exact zeros are omitted because they cannot be represented on logarithmic axes.
For the transformed $f$-mode quantities, the exterior contribution is the largest of the three propagated contributions but remains below a few $10^{-5}$ at the production resolution.

The slow- and rapid-conversion curves for the transformed $f$-mode variables partially overlap, particularly in the TOV and exterior scans.
This is consistent with their common equilibrium background and the weak sensitivity of the $f$-mode frequencies to the interface condition found in the main text.
Here, the overlap indicates similar numerical sensitivities, with each conversion limit evaluated relative to its own finest-step reference solution.

Combining the three step-size scans, the largest resolution-dependent fractional difference in the $p_1$-mode frequencies near the production settings is at the level of a few $10^{-4}$. The cutoff test independently gives relative frequency changes below a few $10^{-4}$ even when the surface cutoff is increased by four orders of magnitude.
These tests indicate that the reported $p_1$-mode trends are not driven by the chosen numerical resolutions or by the surface cutoff.
Moreover, for the representative model tested here, the propagated step-size sensitivity of the $f$-mode universal-relation variables remains below a few $10^{-5}$ near the production settings. This is below the typical $10^{-3}$--$10^{-2}$ averaged-deviation scales that form the main structure of the hybrid-EOS maps discussed in \cref{sec:hybrid_fc_deviation}.
The production resolutions and the choice $P_{\rm cut}/P_c=10^{-14}$ therefore provide adequate numerical stability for the physical comparisons made in this work.

\section{Detailed numerical implementation}
\label{App:numericsprocedures}

This appendix summarizes the numerical procedure used to compute the polar quasi-normal modes. The calculation follows the standard interior--exterior matching strategy for relativistic stellar oscillations: one constructs an interior solution satisfying the center and surface conditions (\cref{sec:BC at C and S}), matches it to the exterior Zerilli solution, and selects the eigenfrequency by imposing the purely outgoing-wave condition at infinity~\cite{Andersson:1995wu}.
A schematic overview of a similar computational workflow is given in Fig.~1 of Ref.~\cite{Kumar:2026tgi}.

\subsection{Interior perturbation equations}
\label{app:interior_equations}
We begin by writing the explicit first-order equations for the interior perturbation variables ${\mathbf{Y}}=(H_1,K,W,X)^T$ used in the numerical implementation. With $\kappa_{\ell}=(\ell-1)(\ell+2)/2$, they are
\begin{align}
H_1' &=
\left[
\frac{1}{2}\left(\lambda'-\nu'\right)
-\frac{\ell+1}{r}
\right]H_1
+\frac{e^\lambda}{r}
\left[
H_0+K-16\pi(\varepsilon+P)V
\right],\label{eq:first_order_systemH1}
\\
K' &=
\frac{1}{r}H_0
+\frac{\kappa_{\ell}+1}{r}H_1
+\left(
\frac{1}{2}\nu'
-\frac{\ell+1}{r}
\right)K
-\frac{8\pi(\varepsilon+P)}{r}e^{\lambda/2}W,\label{eq:first_order_systemK}
\\
W' &=
r e^{\lambda/2}
\left[
\frac{\varepsilon+P}{\Gamma_1 P}
\left(
e^{-\nu}\omega^2 V+\frac{1}{2}H_0
\right)
-\frac{2(\kappa_{\ell}+1)}{r^2}V
+\frac{1}{2}H_0
+K
\right]\nonumber\\
&\quad-\left(
\frac{\ell+1}{r}
+\frac{P'}{\Gamma_1 P}
\right)W \label{eq:first_order_systemW}\\
X' &=
-\frac{\ell}{r}X
+\frac{1}{2}(\varepsilon+P)e^{\nu/2}
\Bigg\{
\left(
\frac{1}{r}
-\frac{1}{2}\nu'
\right)H_0\nonumber\\
&\quad+\left[
r\omega^2 e^{-\nu}
+\frac{\kappa_{\ell}+1}{r}
\right]H_1
+\left(
\frac{3}{2}\nu'
-\frac{1}{r}
\right)K-\nu'\frac{\ell(\ell+1)}{r^2}V
\nonumber\\
&\quad
-\frac{1}{r}
\left[
8\pi(\varepsilon+P)e^{\lambda/2}
+2\omega^2 e^{\lambda/2-\nu}
-r^2\left(r^{-2}e^{-\lambda/2}\nu'\right)'
\right]W
\Bigg\}.
\label{eq:first_order_systemX}
\end{align}
Here primes denote derivatives with respect to $r$, and the algebraic
relations for $H_0$ and $V$ are those given in~\cref{eq:H0_constraint,eq:V_algebraic}. The frequency $\omega$ in these equations is the complex angular frequency defined in Sec.~\ref{sec:polar_perturbations}.

We then describe how this interior system is solved and matched to the exterior spacetime to determine the quasi-normal-mode frequencies.

\subsection{Interior solution}
\label{App:interior}

For a given EOS and central pressure (or energy density), we first solve the TOV equations to obtain the equilibrium background quantities
$m(r)$, $P(r)$, $\varepsilon(r)$, $\nu(r)$, and $\lambda(r)$.
The formal stellar surface $R$ is defined by $P(R)=0$, but in the numerical background construction, we introduce the pressure cutoff
\begin{equation}
    P_{\rm cut}=10^{-14}P_c .
\end{equation}
Once two consecutive TOV integration points bracket $P_{\rm cut}$, the cutoff radius $R_{\rm cut}$ is located by interpolation in $P$, and is used as the numerical approximation to the formal zero-pressure radius $R$. The other background quantities are interpolated to this radius.
To avoid overburdening the notation, in the numerical matching and exterior equations below we suppress the subscript and write
$R\equiv R_{\rm cut}$.

For each trial complex frequency $\omega$, we solve the first-order interior system introduced in \cref{eq:first_order_systemH1,eq:first_order_systemK,eq:first_order_systemW,eq:first_order_systemX} for the perturbation vector $\mathbf{Y}$.
The system is treated numerically as a two-sided boundary-value problem.
Regularity at the center leaves two independent nonsingular solutions (See \cref{sec:BC at C and S}).
We initialize these solutions, denoted by $\mathbf{Y}_4$ and $\mathbf{Y}_5$, at a small radius $r_0>0$ using the regular center expansion and integrate them outward.
At the outer boundary, \cref{eq:surface_condition} leaves three independent surface-compatible solutions, denoted by $\mathbf{Y}_1$, $\mathbf{Y}_2$, and $\mathbf{Y}_3$.
They are initialized at a near-surface interior point $R_G<R_{\rm cut}$ and integrated inward, which will be discussed later.

Then, the five basis solutions are propagated to a common matching radius
$r_m$.
The solution on each side of the interface is written as
\begin{align}
    \mathbf{Y}^{+}(r_m)
    &=
    \eta_1\mathbf{Y}_1(r_m^{+})
    +\eta_2\mathbf{Y}_2(r_m^{+})
    +\eta_3\mathbf{Y}_3(r_m^{+}),
    \\
    \mathbf{Y}^{-}(r_m)
    &=
    \eta_4\mathbf{Y}_4(r_m^{-})
    +\eta_5\mathbf{Y}_5(r_m^{-}),
\end{align}
where $+$ and $-$ denote the outer and the inner sides of the matching interface, respectively.
For a hadronic star, $r_m$ may be chosen at any regular interior radius, and both integrations are matched by continuity of the full perturbation vector,
\begin{equation}
    [\mathbf{Y}]_{r_m}=0 .
\label{eq:regular_interior_matching}
\end{equation}
The resulting eigenfrequency is independent of the choice of $r_m$ up to numerical error.
However, for a hybrid star with a first-order phase transition, we set $r_m=r_t$, so that the two families of solutions are matched directly across the phase interface.
The coefficients $\eta_i$ are determined by imposing the four interface junction conditions summarized in \cref{tab:junction_conditions}.
For a trial frequency $\omega$, the corresponding junction residuals
are
\begin{equation}
    \mathbf{R}_\alpha(\omega,\boldsymbol{\eta})
    \equiv
    \begin{pmatrix}
        H_1^+-H_1^-\\
        K^+-K^-\\
        \Delta P^+-\Delta P^-\\
        \left(\xi^r-\alpha\Delta P/P'\right)^+
        -
        \left(\xi^r-\alpha\Delta P/P'\right)^-
    \end{pmatrix}_{r=r_t}
    =0 .
\end{equation}
Here $\Delta P=-e^{-\nu/2}r^\ell X$, and $P'$ denotes the background pressure gradient on each side of the interface.
As the overall normalization of the perturbation is arbitrary, we fix one coefficient, which may be chosen to be $\eta_5=1$.
Then, the four junction conditions form a $4\times4$ linear system for the remaining coefficients.
The overall numerical procedure for the interior is illustrated in \cref{fig:numerical_matching_scheme}.

We now describe the near-surface construction of the three surface-compatible basis solutions.
Because their inward integration begins at $R_G<R_{\rm cut}$ rather than at the formal surface itself, directly imposing $X(R_G)=0$ would apply the fluid boundary condition at the wrong radius.
At the chosen near-surface starting point $R_G$, we approximate the remaining outer layer by a local polytrope with the EOS-determined index
\begin{equation}
    N_G
    =
    \frac{1}{\Gamma_1(R_G)-1}.
\end{equation}
Thus, $N_G$ is fixed locally by the background EOS and is not an additional fitting parameter. 
Within the local-polytropic approximation, the leading regular surface behavior is
\begin{equation}
    X(r)\propto (R-r)^{N_G+1},
    \label{eq:leading regular surface behavior}
\end{equation}
which gives
\begin{equation}
    X(R_G)
    =
    \frac{R_G-R}{N_G+1}X'(R_G)
    \simeq
    \frac{R_G-R_{\rm cut}}{N_G+1}X'(R_G).
\label{eq:surface_correction}
\end{equation}
Note that $R$ denotes the formal zero-pressure radius in \cref{eq:leading regular surface behavior,eq:surface_correction} before the approximation $R\simeq R_{\rm cut}$ is made.
The second form is the one used in the numerical implementation, with $R_{\rm cut}$ serving as the approximation to the formal zero-pressure surface.

After the matching coefficients have been determined, the physical interior solution is reconstructed from the two basis families up to $R_G$.
Across the thin interval between $R_G$ and $R_{\rm cut}$, the matched perturbation state is evaluated by first-order extrapolation from $R_G$.
The resulting values of $H_1(R_{\rm cut})$ and $K(R_{\rm cut})$ are then used to construct the surface Zerilli data [see Eq.~(A8) in Ref.~\cite{Zhao:2022tcw}] required for matching to the exterior solution, as discussed below.

\subsection{Exterior solution and outgoing condition}
\label{App:exterior}
The exterior calculation follows the complex-coordinate phase-amplitude method of Andersson, Kokkotas, and Schutz (AKS)~\cite{Andersson:1995wu}. We adopt the same harmonic time dependence $e^{+i\omega t}$ as the AKS formulation. 
The simplified overall exterior matching procedure is summarized schematically
in \cref{fig:surface_mode_condition}.

Outside the star, the fluid perturbations vanish, and the metric perturbations are described by the Zerilli function $Z(r)$, which satisfies
\begin{equation}
    \left[
    \frac{{\rm d}^2}{{\rm d}r_*^2}
    +\omega^2
    -V_Z(r)
    \right]Z(r)=0 ,
\label{eq:zerilli_exterior}
\end{equation}
where $V_Z(r)$ is the Zerilli potential~\cite{Zerilli:1970se}
[for the convention used here, see Eq.~(12) of~\cite{Andersson:1995wu} or Eq.~(A36) of~\cite{Lindblom:1983ps}], and $r_*$ is the tortoise coordinate defined by
\begin{equation}
    \frac{{\rm d}}{{\rm d}r_*}
    =
    \left(1-\frac{2M}{r}\right)
    \frac{{\rm d}}{{\rm d}r}.
\end{equation}
In our convention, the asymptotic solution takes the form
\begin{equation}
Z
\sim
A_{\rm out}e^{-i\omega r_*}
+
A_{\rm in}e^{+i\omega r_*}.
\label{eq:zerilli_asymptotic}
\end{equation}
Together with the time dependence $e^{+i\omega t}$, the first term is proportional to $e^{i\omega(t-r_*)}$ and is therefore outgoing, whereas the second term is ingoing.
The physical interior solution supplies the metric perturbations at
the numerical stellar surface.
These are converted to the Zerilli surface data
\begin{equation}
    Z_s=Z(R),
    \qquad
    Z_s'
    =
    \left.
    \frac{{\rm d}Z}{{\rm d}r_*}
    \right|_R ,
\label{eq:zerilli_surface_data}
\end{equation}
using Eq.~(A8) of Ref.~\cite{Zhao:2022tcw}
\begin{equation}
\begin{pmatrix}
K\\
H_1
\end{pmatrix}
=
\begin{pmatrix}
g & 1\\
h & k
\end{pmatrix}
\begin{pmatrix}
Z/R\\
{\rm d}Z/{\rm d}r_*
\end{pmatrix},
\label{eq:KH1_Zerilli_transform}
\end{equation}
where
\begin{align}
g
&=
\frac{
\kappa_{\ell}(\kappa_{\ell}+1)+3\kappa_{\ell} b+6b^2
}{
\kappa_{\ell}+3b
},\\
h
&=
\frac{
\kappa_{\ell}-3\kappa_{\ell} b-3b^2
}{
(1-2b)(\kappa_{\ell}+3b)
},\\
k
&=
\frac{1}{1-2b},
\qquad
b=\frac{M}{R}.
\end{align}
Here and below, $R$ denotes the numerical surface $R_{\rm cut}$, as
specified in \cref{App:interior}.
The exterior calculation determines whether these surface data are
compatible with a purely outgoing solution at infinity.

For the exterior AKS problem, we define
\begin{equation}
    f(r)=1-\frac{2M}{r},
    \qquad
    Z=f^{-1/2}\Psi,
\end{equation}
and introduce the dimensionless variables
\begin{equation}
    z=\frac{r}{M},
    \qquad
    \Omega=M\omega.
\end{equation}
The transformed Zerilli equation can then be written as
\begin{equation}
    \frac{{\rm d}^2\Psi}{{\rm d}z^2}
    +
    \mathcal{U}(z,\Omega)\Psi
    =0 ,
\label{eq:aks_transformed_wave_equation}
\end{equation}
where 
\begin{equation}
    \mathcal{U}(z,\Omega)
    \equiv
    M^2U(Mz,\omega)
\end{equation}
is the dimensionless form of the effective potential $U(r,\omega)$, defined in Eq.~(16) of Ref.~\cite{Andersson:1995wu}.

Let $q_{\rm AKS}(r)$ denote the dimensional phase-amplitude function
introduced as $q(r)$ in Eqs.~(17)--(18) of
Ref.~\cite{Andersson:1995wu}, and define its dimensionless counterpart
as
\begin{equation}
    q(z)\equiv Mq_{\rm AKS}(Mz).
\end{equation}
The two phase-amplitude solutions can then be written, up to an overall
normalization, as
\begin{equation}
    \Psi^\pm
    =
    q^{-1/2}
    \exp\left[
        \pm i\int q\,{\rm d}z
    \right].
\label{eq:phase_amplitude_solutions}
\end{equation}
In the AKS convention, $\Psi^+$ and $\Psi^-$ represent the incoming
and outgoing branches, respectively.
The dimensionless phase function satisfies
\begin{equation}
    \frac{1}{2q}\frac{{\rm d}^2q}{{\rm d}z^2}
    -
    \frac{3}{4q^2}
    \left(
        \frac{{\rm d}q}{{\rm d}z}
    \right)^2
    +
    q^2
    -
    \mathcal{U}(z,\Omega)
    =0 .
\label{eq:dimensionless_phase_equation}
\end{equation}

For each complex trial frequency, \cref{eq:dimensionless_phase_equation} is integrated inward from an asymptotic point to $z_R={R}/{M}$ along the straight contour
\begin{equation}
    z(\rho)=z_R+\rho e^{i\theta},
    \qquad
    \theta=-\arg\Omega.
\label{eq:aks_complex_contour}
\end{equation}
This contour is parallel to the asymptotic anti-Stokes direction and suppresses the exponential contamination that would arise from integrating a quasi-normal-mode solution along the real radial axis.
For real trial frequencies, $\theta=0$, and the contour reduces to the
real axis.

At the outer endpoint, we use the leading-WKB data
\begin{equation}
    q\simeq\sqrt{\mathcal{U}},
    \qquad
    \frac{{\rm d}q}{{\rm d}z}
    \simeq
    \frac{1}{2q}
    \frac{{\rm d}\mathcal{U}}{{\rm d}z},
\label{eq:aks_outer_data}
\end{equation}
choosing the square-root branch such that
$q\rightarrow\Omega$ in the asymptotic region.
The inward integration transports the asymptotic wave basis to the
stellar surface and yields
\begin{equation}
    q_R=q(z_R),
    \qquad
    q_{R,z}
    =
    \left.
    \frac{{\rm d}q}{{\rm d}z}
    \right|_{z_R}.
\label{eq:q_surface_data}
\end{equation}
The outer endpoint $z_{\rm out}=z_R+\rho_{\rm out}e^{i\theta}$ is chosen deep in the asymptotic region, with $\rho_{\rm out}$ typically of order $10^3$ or larger and increased further for small $|\Omega|$; once this scale is reached, further increasing $\rho_{\rm out}$ leaves the surface phase data $(q_R,q_{R,z})$ and the extracted quasi-normal-mode frequency unchanged within numerical accuracy.

To connect the dimensionless surface phase data in \cref{eq:q_surface_data} with the dimensional quantities entering Eq.~(24) of Ref.~\cite{Andersson:1995wu}, we note that
\begin{equation}
    q_{\rm AKS}(R)
    =
    \frac{q_R}{M},
    \qquad
    \left.
    \frac{{\rm d}q_{\rm AKS}}{{\rm d}r}
    \right|_R
    =
    \frac{q_{R,z}}{M^2}.
\label{eq:q_surface_conversion}
\end{equation}
Since $Z=f^{-1/2}\Psi$, the surface data used in the exterior AKS problem satisfy
\begin{equation}
    \Psi_R=\sqrt{f_R}\,Z_s,
    \qquad
    \left.
\frac{{\rm d}\Psi}{{\rm d}r}
\right|_R
=
\frac{1}{\sqrt{f_R}}
\left[
    Z_s'
    +
    \frac{M}{R^2}Z_s
\right],
\label{eq:zerilli_psi_surface_conversion}
\end{equation}
where $f_R=1-{2M}/{R}$ is the Schwarzschild factor.

Substituting \cref{eq:q_surface_conversion} into Eq.~(24) of Ref.~\cite{Andersson:1995wu}, the incoming amplitude in the AKS time convention is
\begin{align}
A_{\rm in}^{\rm AKS}
&=
-\frac{i}{2}
\sqrt{\frac{M}{q_R}}\,
f_R^{-1/2} \mathcal D_{\rm in}(\omega),\\
\mathcal D_{\rm in}(\omega)
&=
Z_s
\left[
\frac{f_R}{M}
\left(
    iq_R+\frac{q_{R,z}}{2q_R}
\right)
+
\frac{M}{R^2}
\right]
+
Z_s'.
\label{eq:Ain_AKS_dimensionless}
\end{align}
The quasi-normal-mode condition is then
\begin{equation}
A_{\rm in}^{\rm AKS}(\omega)=0.
\label{eq:outgoing-wave-condition}
\end{equation}
For numerical root finding, we solve the equivalent condition $\mathcal D_{\rm in}(\omega)=0$. After normalizing the reconstructed eigenfunction according to
\cref{eq:mode_normalization}, we define the accepted-root residual as $ \left|\mathcal D_{\rm in}(\omega)\right|$, whose magnitude is independent of the arbitrary overall complex amplitude of the unnormalized eigenfunction. For the quadrupolar calculation, with all lengths expressed in km and with the eigenfunction normalization specified above, an accepted root satisfies $\left|\mathcal D_{\rm in}(\omega)\right|<10^{-8} \, {\rm km}^{-2}$.

\subsection{Mode search}
\label{App:Mode search}

This subsection describes the numerical search for complex roots, the convention used for the eigenfunctions, and the identification and tracking of the $f$- and $p_1$-mode branches.
Candidate modes are identified by scanning the outgoing-wave residual over a real-frequency grid, and the corresponding complex frequencies are refined using Muller's method.

Because a quasi-normal-mode eigenfunction is complex and is determined only up to an arbitrary nonzero complex multiplicative constant, we first fix its overall magnitude using the following displacement norm
\begin{equation}
    \mathcal{A}
    =
    \left\{
    \frac{1}{M R^\ell}
    \int_0^R
    (\varepsilon+P)
    e^{\frac{\nu+\lambda}{2}}
    r^{2\ell}
    \left[
        |W(r)|^2
        +
        \ell(\ell+1)|V(r)|^2
    \right]
    {\rm d}r
    \right\}^{\frac{1}{2}}.
\label{eq:mode_normalization}
\end{equation}
The full reconstructed perturbation solution is divided by $\mathcal{A}$, so that the dimensionless norm in \cref{eq:mode_normalization} is unity. This fixes the overall magnitude of the eigenfunction, but leaves its common complex phase arbitrary. 
For the purpose of counting radial nodes, we fix this phase by applying a common phase rotation, so that the first resolved nonzero value of $W$ outside the central region is real and positive. 
Nodes are then identified from sign changes of the real part of the phase-fixed $W(r)$ profile. 
A sign reversal between the two one-sided interface values is also counted as an interface-crossing node, even when neither one-sided value vanishes exactly at $r=r_t$. 
This interface-node convention records a reversal of the radial fluid motion across the phase boundary.

With this normalization and phase convention, mode identification proceeds in stages. 
For each stellar model, we first search a prescribed low-frequency interval expected to contain the $f$-mode.
The radial-displacement eigenfunction $W(r)$ of a candidate root should have no internal nodes, consistent with the standard zero-node characterization of the hadronic $f$-mode. 
Also, it should be tracked continuously as the central pressure is varied along each fixed-EOS sequence. 
Any lower-frequency $g$- or interface-mode roots that enter the search interval are identified from their eigenfunction morphology and excluded from the present analysis.

After the $f$-mode branch has been identified, we search the higher-frequency spectrum for pressure modes.
For simple acoustic modes, the radial order is commonly associated with the number of internal nodes of the radial displacement.
In the standing-wave picture, these nodes arise from interference between oppositely propagating acoustic components, with the spatial pattern determined by wave propagation and the boundary conditions.
However, the total radial-node count is not a general mode-ordering rule for nonradial stellar oscillations.
As discussed in \cite[Sec.~3.5.2]{Aerts:2010},
the phase-diagram classification of nonradial stellar oscillations assigns opposite contributions to acoustic- and gravity-type nodes.
Within this classification, additional zeros of the radial displacement can appear in pairs without changing the mode order.
Explicit examples are given in \cite[Table~4 and Fig.~9]{Osaki_1975}.

\begin{figure}[t]
    \centering
    \includegraphics[width=1\linewidth]{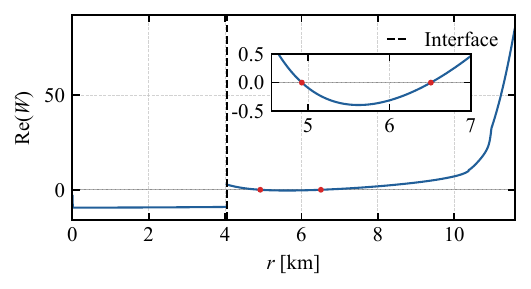}
    \caption{Real part of the normalized, dimensionless radial-displacement eigenfunction $W$ for the $p_1$-mode listed in \cref{tab:mode_nodes_example}. The dashed vertical line marks the phase interface, across which $W$ changes sign discontinuously. Red dots mark two additional nodes in the continuous part of the profile, shown in greater detail in the inset.}
    \label{fig:p1_mode_W_profile}
\end{figure}

In our hybrid-star models, the sharp phase boundary introduces additional matching conditions that govern reflection and transmission at the interface and modify the standing-wave pattern. The radial-displacement profile may reverse sign across the interface through a finite jump, while an additional pair of smooth zero crossings may appear away from the interface. Consequently, the number of nodes in the radial displacement can exceed the radial mode order. 
We illustrate this point with the representative rapid-conversion model in \cref{tab:mode_nodes_example}. We use $p_1$ to denote the lowest-frequency acoustic branch above the identified $f$-mode, with the assignment based on the frequency spectrum, the eigenfunctions, and branch continuation. 
\Cref{fig:p1_mode_W_profile} shows the radial-displacement eigenfunction $W$ of the $p_1$-mode listed in \cref{tab:mode_nodes_example}. In addition to the discontinuous sign reversal at the phase interface, the profile contains two additional nodes away from the interface. This behavior is not restricted to the $p_1$-mode: some of the $p_2$-modes in our models likewise exhibit an additional pair of nodes, bringing their total count to four. 

\begin{table}[t]
\centering
\caption{
Mode frequencies and $\mathrm{Re}(W)$-node counts for the rapid-conversion SLy4 hybrid star, with ($p_t$, $\Delta\varepsilon$) = (34.079, 147.716)~$\mathrm{MeV\,fm^{-3}}$, and $c_{s,\rm con}^2=1$.
The stellar background has $P_c=53.415\,\mathrm{MeV\,fm^{-3}}$, $M=1.015\,M_\odot$, and $R=11.594\,\mathrm{km}$.
Node counts include sign changes at the phase interface.
}
\label{tab:mode_nodes_example}
\renewcommand{\arraystretch}{1.3}
\setlength{\tabcolsep}{12pt}
\begin{tabular}{c c c}
\hline\hline
$\omega_{\rm re}/(2\pi)\,[\mathrm{kHz}]$
& $\mathrm{Re}(W)$ nodes & Mode \\
\hline
1.81132  & 0 & $f$   \\
5.79353  & 3 & $p_1$ \\
7.27175  & 2 & $p_2$ \\
8.09096  & 3 & $p_3$ \\
9.43400  & 4 & $p_4$ \\
11.66368 & 5 & $p_5$ \\
12.96245 & 6 & $p_6$ \\
\hline\hline
\end{tabular}
\end{table}

Over the parameter ranges shown in \cref{fig:eos_parameter_space_SLY4}, the selected $f$- and $p_1$-mode branches remain well separated in the complex-frequency plane, and we find no indication of an interchange or avoided crossing between these two branches.

\end{document}